\documentclass[12pt]{article}
\usepackage{amssymb,amscd,array}
\catcode `\@=11
\@addtoreset{equation}{section}

\newtheorem{thm}{Theorem}[section]
\newtheorem{rem}[thm]{Remark}

\newtheorem{prop}[thm]{Proposition}

\def\qed{\blacksquare}
\newcommand{\be}{\begin{equation}}
\newcommand{\ee}{\end{equation}}
\newcommand{\bea}{\begin{eqnarray}}
\newcommand{\eea}{\end{eqnarray}}

\newcommand{\N}{\mathbb{N}}
\newcommand{\C}{\mathbb{C}}
\def\d{\partial}
\begin{document}
\begin{titlepage}

\begin{center}
{\bf \Large{Anomaly-Free Gauge Models: A Causal Approach\\}}
\end{center}
\vskip 1.0truecm
\centerline{D. R. Grigore, 
\footnote{e-mail: grigore@theory.nipne.ro}}
\vskip5mm
\centerline{Department of Theoretical Physics}
\centerline{Institute for Physics and Nuclear Engineering ``Horia Hulubei"}
\centerline{Bucharest-M\u agurele, P. O. Box MG 6, ROM\^ANIA}

\vskip 2cm
\bigskip \nopagebreak
\begin{abstract}
\noindent
The gauge invariance of some massless Yang-Mills models can be proved for a large class of groups using Polchinski flow equations 
approach. In this paper we provide an alternative proof based on the causal approach.
The proof is purely algebraic and is based on the analysis of the anomalies. More precisely, one 
can prove that the anomalies are verifying some consistency equations of Wess-Zumino type. In the 
massless $SU(2)$ Yang-Mills case, this is enough to prove that they are absent. The same is true for QED.
\end{abstract}
Contribution to the conference ``Quantum Fields and Nonlinear Phenomena"
18-22 April 2018, Sinaia, Romania
\end{titlepage}

\section{Introduction}

The general framework of perturbation theory consists in the construction of 
the chronological products such that Bogoliubov axioms are verified 
\cite{BS}, \cite{EG}, \cite{DF}, \cite{Gl}, \cite{Sc2}, \cite{Sto1}; for every set of Wick monomials 
$ 
A_{1}(x_{1}),\dots,A_{n}(x_{n}) 
$
acting in some Fock space
$
{\cal H}
$
generated by the free fields of the model, one associates the operator
$ 
T_{A_{1},\dots,A_{n}}(x_{1},\dots,x_{n}); 
$  
all these expressions are in fact distribution-valued operators called chronological products. 
Sometimes it is convenient to use another notation: 
$ 
T(A_{1}(x_{1}),\dots,A_{n}(x_{n})). 
$
These products are constrained by some natural axioms (due to Bogoliubov) and expressing causality, unitarity and 
Poincar\'e covariance. 
The chronological products are not uniquely defined: one can add quasi-local expressions (i.e. distribution-valued
operators with support on the small diagonal
$
x_{1} = \cdots = x_{n}
$)
but there are some natural limitation on the arbitrariness. There are various ways to construct the chronological products. 

(a) In the original proof of Hepp \cite{H} one rewrites Bogoliubov axioms in terms of vacuum averages
$
< \Omega, T_{A_{1},\dots,A_{n}}(x_{1},\dots,x_{n})\Omega>
$
(more precisely the contributions associated to various Feynman graph). One needs a regularization procedure for the 
Feynman amplitudes. Moreover, one proves that the renormalized Feynman amplitudes can be obtained from the formal
Feynman rules if one adds appropriate counterterms in the interaction Lagrangian.

(b) In Polchinski flow equations approach \cite{P}, \cite{S} one considers an ultra-violet cut-off 
$
\Lambda
$
for the Feynman amplitudes and establishes some differential equations (in this parameter) for these amplitudes.
The equations have such a structure that one can obtain the Feynman amplitudes by some recursive procedure and integration
of these differential equations. The computations are usually done in the Euclidean framework and is less obvious that the end 
result will verify Bogoliubov axioms.

(c) The causal approach due to Epstein and Glaser \cite{EG}, \cite{Gl} is a recursive procedure for the basic objects
$ 
T(A_{1}(x_{1}),\dots,A_{n}(x_{n}))
$
and reduces the induction procedure to a distribution splitting of some distributions with causal support.  
In an equivalent way, one can reduce the induction procedure to the process of extension of distributions \cite{PS}. 
An equivalent point of view uses retarded products \cite{St1} instead of chronological products.

Gauge theories are described using {\it ghost fields} which are not physical. Such theories are defined in a Fock space
$
{\cal H}
$
with indefinite metric, generated by physical and un-physical fields, so it contains physical and un-physical states.
A physically reasonable theory should be such that the $S$-matrix (or more precisely the chronological products)
should leave invariant the physical states. Again, there are various ways to achieve this goal.

(A) In BRST approach one can try to make sense of the formal path integral and ends up with some consistency relation - 
the master equation \cite{HT}. Presumably, if a solution of this equation can be found, one would be able to construct 
the chronological products with the desired properties.

(A') A variant of the preceding idea is the use of the Zinn-Justin relation \cite{Z-J}.

(B) One assumes the existence of an operator $Q$ called {\it gauge charge} which verifies
$
Q^{2} = 0
$
and such that the {\it physical Hilbert space} is by definition
$
{\cal H}_{\rm phys} \equiv Ker(Q)/Ran(Q).
$
The graded commutator
$
d_{Q}
$
of the gauge charge with any operator $A$ of fixed ghost number
\be
d_{Q}A = [Q,A]
\ee
is a co-chain operator in the space of Wick polynomials. From now on
$
[\cdot,\cdot]
$
denotes the graded commutator.
 
A gauge theory assumes also that there exists a Wick polynomial of null ghost number
$
T(x)
$
called {\it the interaction Lagrangian} such that
\be
~[Q, T(x)] = i \partial_{\mu}T^{\mu}(x)
\label{gauge-1}
\ee
for some other Wick polynomials
$
T^{\mu}.
$
This relation means that the expression $T$ leaves invariant the physical states, at least in the adiabatic limit. 

In all known models there exists a chain of Wick polynomials
$
T^{\mu},~T^{\mu\nu},~T^{\mu\nu\rho},\dots
$
such that:
\be
~[Q, T] = i \partial_{\mu}T^{\mu}, \quad
[Q, T^{\mu}] = i \partial_{\nu}T^{\mu\nu}, \quad
[Q, T^{\mu\nu}] = i \partial_{\rho}T^{\mu\nu\rho},\dots
\label{descent}
\ee
with
$
T^{\mu\nu},~T^{\mu\nu\rho},\dots
$
completely antisymmetric in all indexes so we can also use a compact notation
$
T^{I}
$
where $I$ is a collection of indexes; when convenient we emphasize by brackets 
the complete antisymmetry in these indexes. One can write compactly the relations (\ref{descent}) as follows:
\be
d_{Q}T^{I} = i~\partial_{\mu}T^{I\mu}.
\label{descent1}
\ee

We can construct the chronological products
$$
T^{I_{1},\dots,I_{n}}(x_{1},\dots,x_{n}) \equiv T(T^{I_{1}}(x_{1}),\dots,T^{I_{n}}(x_{n}))
$$
according to Epstein-Glaser procedure. We say that the theory is gauge invariant in all orders of the perturbation theory 
if the following set of identities generalizing (\ref{descent1}):
\be
d_{Q}T^{I_{1},\dots,I_{n}} = 
i \sum_{l=1}^{n} (-1)^{s_{l}} {\partial\over \partial x^{\mu}_{l}}
T^{I_{1},\dots,I_{l}\mu,\dots,I_{n}}
\label{gauge-n}
\ee
are true for all 
$n \in \N$
and all
$
I_{1}, \dots, I_{n}.
$
Here we have defined
\be
s_{l} \equiv \sum_{j=1}^{l-1} |I|_{j}.
\ee

Such identities can be usually broken by {\it anomalies} i.e. expressions of the type
$
A^{I_{1},\dots,I_{n}}
$
which are quasi-local and might appear in the right-hand side of the relation (\ref{gauge-n}). 

The approach (A') combined with Polchinski method (b) has been used in \cite{FHH} for the case of a massless system of
Yang-Mills fields with a compact simple Lie algebra. As remarked in this reference, this approach avoids the problem of
the infra-red divergences which are treated only formally in the BRST approach (A). The expression of the 
zero-order perturbation theory - the Lagrangian - is an external data coming from classical field theory.

Our approach based on Epstein-Glaser method (c) with gauge invariance in the sense (B) -see (\ref{gauge-n}) above - also
avoids the infra-red divergences. One computes the anomalies in orders one and two of perturbation theory and imposing their 
cancellation one obtains various restrictions on the expression of the interaction Lagrangian. 
The outcome is the well-known expression of the Yang-Mills Lagrangian. 

The anomalies in the second order of perturbation theory can be computed using the formalism of off-shell fields. 
This formalism was used in \cite{EG} to prove the equivalence between the causal approach and the counterterm approach in
renormalization. Using induction we prove that for an massless
$
su(2)
$
Yang-Mills model there are no anomalies in higher orders of perturbation theory. The method we use is based by the version
of the Wess-Zumino consistency relations adapted to the causal approach. We simplify the analysis from \cite{ano} concerning this
aspect of the computations.

In the next Section we remind our definition of free fields. We avoid explicit formulas using the reconstruction theorem of Wightman.
In Section \ref{int} we recall the main result concerning the interaction Lagrangians 
for the most simple model with higher spin fields, namely the pure Yang-Mills fields model. 
In Section \ref{off} we  remind the reader the off-shell formalism \cite{off}. In Section \ref{ano} we determine the structure of 
the anomalies in an
arbitrary order of the perturbation theory and prove the for the algebra
$
su(2)
$
such an anomaly must be null. In the last Section \ref{qed} we use the same methods for quantum electrodynamics. Here 
instead of 
$
su(2)
$
invariance one must use charge conjugation invariance. 

The main point of the above computations is that they are long but elementary. It is a goal to extend such elementary
methods to more general models. 
\newpage

\section{Free Fields\label{free}}

We will adopt the description of free quantum fields given by the reconstruction theorem from axiomatic field theory 
\cite{J}, \cite{SW} based on Borchers algebras. We follow essentially theorem 8.8, pg. 324 of \cite{BLOT}.
In this approach one can construct a quantum field giving the Wightman $n$-points distributions and the statistics. 
For a free field it is sufficient to give the Wightman $2$-points distribution and generate the rest according to Wick theorem. 
This point of view has been advocated many times in the literature: see for instance \cite{Ha} (sect. II.2.2, end of pg. 62)
where it is observed that one should consider the truncated Wightman functions and define a free field by the 
condition that these truncated functions should be null for 
$
n \geq 3.
$
This point of view is especially useful when considering free fields on curved background manifolds. 
In our context, this approach seems to be the most convenient. We use formal distribution notations for simplicity.
In \cite{off} one can find the treatment of a real scalar field in this {\it reconstructive approach.} Here we just
give the relevant formulas the the Yang-Mills case. 

The generic case is the massless vector field. In this case we consider the vector space 
$
{\cal H}
$
of Fock type generated (in the sense of Borchers theorem) by the following fields:
$
(v^{\mu}, u, \tilde{u})
$
where the non-zero $2$-point distributions are
\bea
<\Omega, v^{\mu}(x_{1}) v^{\nu}(x_{2})\Omega> = 
i~\eta^{\mu\nu}~D_{0}^{(+)}(x_{1} - x_{2}),
\nonumber \\
<\Omega, u(x_{1}) \tilde{u}(x_{2})\Omega> = - i~D_{0}^{(+)}(x_{1} - x_{2}),
\qquad
<\Omega, \tilde{u}(x_{1}) u(x_{2})\Omega> = i~D_{0}^{(+)}(x_{1} - x_{2}).
\label{2-massless-vector}
\eea

We remark that the form
$
<\cdot, \cdot>
$
defined above cannot be positively defined, but it is sesquilinear. We also assume the following self-adjointness properties:
\be
v_{\mu}^{\dagger} = v_{\mu}, \qquad 
u^{\dagger} = u, \qquad
\tilde{u}^{\dagger} = - \tilde{u}
\label{adj-vector-null}
\ee
and that the field 
$
v^{\mu}
$
is Bose and the fields
$
u, \tilde{u}
$
are Fermi. We generate the $n$-point functions such that the truncated Wightman functions are null for 
$
n \geq 2.
$
When defining the representation of the Lorentz group we consider that the first field is vector and the last two are scalars. 
Because of the ``wrong" statistics the sesquilinear form is not positively defined. Nevertheless, because it is non-degenerated, 
we can prove that we have Klein-Gordon equations of null mass:
\be
\square~v^{\mu} = 0 \qquad \square u = 0 \qquad \square \tilde{u} = 0
\label{KG-vector-null-eq}
\ee
and the canonical commutations relation:
\bea
[v^{\mu}(x_{1}), v^{\nu}(x_{2}) ] = i~\eta^{\mu\nu}~D_{0}(x_{1} - x_{2})
\nonumber \\
\{ u(x_{1}), \tilde{u}(x_{2}) \} = - i~D_{0}(x_{1} - x_{2})
\label{CCR-vector-null}
\eea
and all other (anti)commutators are null.

We can obtain a {\it bona fid\ae} scalar product introducing the so-called {\it gauge charge} i.e. an operator $Q$ defined by:
\bea
~[Q, v^{\mu}] = i~\partial^{\mu}u,\qquad
\{ Q, u \} = 0,\qquad
\{Q, \tilde{u}\} = - i~\partial_{\mu}v^{\mu}
\nonumber \\
Q \Omega = 0.
\label{Q-vector-null}
\eea
Using these relation one can compute the action of $Q$ on any state generated by a polynomial in the fields applied on 
the vacuum by commuting the operator $Q$ till it hits the vacuum and gives zero. However, because of the canonical 
commutation relations the writing of a polynomial state is not unique. One can prove that the operator $Q$ leaves invariant 
the canonical (anti)commutation relations given above and this leads to the consistency of the definition. 
Then one shows that the operator $Q$ squares to zero:
\be
Q^{2} = 0
\ee 
and that the factor space
$
Ker(Q)/Ran(Q)
$
is isomorphic to the Fock space particles of zero mass and helicity $1$ (photons and gluons) \cite{cohomology}.

We can generalize this case considering the tensor product of $r$ copies of massless vector fields, i.e. we consider the set of fields 
$
(v^{\mu}_{a}, u_{a}, \tilde{u}_{a}),~a = 1,\dots,r
$
of null mass and we extend in an obvious way the definitions of the scalar product and of the gauge charge.
\newpage
\section{Interactions\label{int}}

The discussion from the Introduction provides the physical justification for determining the cohomology of the operator 
$
d_{Q} = [Q,\cdot]
$
induced by $Q$ in the space of Wick polynomials. A polynomial 
$
p \in {\cal P}
$
verifying the relation
\be
d_{Q}p = i~\d_{\mu}p^{\mu}
\label{rel-co}
\ee
for some polynomials
$
p^{\mu}
$
is called a {\it relative co-cycle} for 
$
d_{Q}.
$
The expressions of the type
\be
p = d_{Q}b + i~\d_{\mu}b^{\mu}, \qquad (b, b^{\mu} \in {\cal P})
\ee
are relative co-cycles and are called {\it relative co-boundaries}. We denote by
$
Z_{Q}^{\rm rel}, B_{Q}^{\rm rel} 
$
and
$
H_{Q}^{\rm rel}
$
the corresponding cohomological spaces. In (\ref{rel-co}) the expressions
$
p_{\mu}
$
are not unique. It is possible to choose them Lorentz covariant. We have a general description of the most 
general form of the interaction of the previous fields \cite{cohomology}. Summation over the dummy indexes is used everywhere. 
For simplicity we do not write the double dots of the Wick product notations.

\begin{thm}
Let $T$ be a relative co-cycle in the variables 
$
(v^{\mu}_{a}, u_{a}, \tilde{u}_{a}),~a = 1,\dots,r
$
which is tri-linear in the fields, of canonical dimension
$
\omega(T) \leq 4
$
and ghost number
$
gh(T) = 0.
$
Then:
(i) $T$ is (relatively) cohomologous to a non-trivial co-cycle of the form:
\be
t = f_{abc} \left( {1\over 2}~v_{a\mu}~v_{b\nu}~F_{c}^{\nu\mu}
+ u_{a}~v_{b}^{\mu}~\d_{\mu}\tilde{u}_{c}\right)
\label{int-ym}
\ee

(ii) The relation 
$
d_{Q}t = i~\d_{\mu}t^{\mu}
$
is verified by:
\be
t^{\mu} = f_{abc} \left( u_{a}~v_{b\nu}~F^{\nu\mu}_{c} -
{1\over 2} u_{a}~u_{b}~\d^{\mu}\tilde{u}_{c} \right)
\label{t-mu-ym}
\ee

(iii) The relation 
$
d_{Q}t^{\mu} = i~\d_{\nu}t^{\mu\nu}
$
is verified by
\be
t^{\mu\nu} \equiv {1\over 2} f_{abc}~u_{a}~u_{b}~F_{c}^{\mu\nu}.
\label{t-munu-ym}
\ee
and we have
$
d_{Q}t^{\mu\nu} = 0.
$

(iv) The constants
$
f_{abc}
$
must be completely antisymmetric
\be
f_{abc} = f_{[abc]}
\label{anti-f}
\ee
and the expressions given above are self-adjoint iff the constants
$
f_{abc}
$
are real. Here we have defined the gauge invariants which are not coboundaries
\be
F^{\mu\nu}_{a} \equiv \d^{\mu}v^{\nu}_{a} - \d^{\nu}v^{\mu}_{a}, 
\quad \forall a = 1,\dots,r
\ee 
\label{t-ym}
\end{thm}

There are different ways to obtain the preceding results. One can proceed by brute force, making an ansatz for the expressions
$
T^{I}
$
and solving the identities of the type (\ref{descent1}) as it is done in \cite{Sc2}. 
There are some tricks to simplify such a computation. The first one makes an ansatz for 
$
T
$
and eliminates the most general relative cocycle. Then one computes 
$
d_{Q}T
$
and writes it as a total divergence plus terms without derivatives on the ghost fields. 
Another trick is to use the so-called descent procedure. We briefly present the first line of proof. It starts from the general form:
\bea
T = f^{(1)}_{abc} v_{a}^{\mu} v_{b}^{\nu} \partial_{\mu}v_{c\mu} 
+ f^{(2)}_{abc} v_{a}^{\mu} v_{b\mu} \partial_{\nu}v_{c}^{\nu} 
+ f^{(3)}_{abc}~\epsilon_{\mu\nu\rho\sigma}~v_{a}^{\mu} v_{b}^{\nu} \partial^{\sigma}v_{c}^{\rho}
\nonumber \\
+ g^{(1)}_{abc} v^{\mu}_{a} u_{b} \partial_{\mu}\tilde{u}_{c}
+ g^{(2)}_{abc} \partial_{\mu}v^{\mu}_{a} u_{b} \tilde{u}_{c}
+ g^{(3)}_{abc} v^{\mu}_{a} \partial_{\mu}u_{b} \tilde{u}_{c}.
\eea
Eliminating relative coboundaries we can fix: 
\be
f^{(1)}_{abc} = - f^{(1)}_{bac},\qquad
f^{(2)}_{abc} = 0,\qquad
g^{(3)}_{abc} = 0,\qquad
g^{(2)}_{abc} = g^{(2)}_{bac}.
\ee
Then we obtain easily:
\be
d_{Q}T = i u_{a} T_{a}  + {\rm total~div}
\ee
where:
\bea
T_{a} = - 2 f^{(1)}_{abc}~\partial^{\nu}v^{\mu}_{b}~\partial_{\mu}v_{c\nu}
+ (f^{(1)}_{cba} + g^{(2)}_{bac})~
\partial_{\mu}v^{\mu}_{b}~\partial_{\nu}v_{c}^{\nu}
\nonumber \\
+ (- f^{(1)}_{abc} + f^{(1)}_{cba} + f^{(1)}_{bca} + g^{(1)}_{bca})~
v^{\mu}_{b}~\partial_{\mu}\partial_{\nu}v^{\nu}_{c}
\nonumber \\
- 2 f^{(3)}_{abc}~\epsilon_{\mu\nu\rho\sigma}
~\partial^{\mu}v^{\nu}_{b}~\partial_{\sigma}v_{c\rho}.
\eea

Now the gauge invariance condition (\ref{gauge-1}) becomes
\be
u_{a} T_{a} = \partial_{\mu}t^{\mu}
\label{gauge-11}
\ee
for some expression
$
t^{\mu}
$
which has, from power counting arguments, the general form
\be
t^{\mu} = u_{a}~t^{\mu}_{a} + \partial^{\mu}u_{a}~t_{a}
+ \partial_{\nu}u_{a}~t^{\mu\nu}_{a}
\ee
where the polynomial
$
t^{\mu\nu}_{a}
$
does not contain terms with the factor 
$
\eta^{\mu\nu}.
$
Then the relation (\ref{gauge-11}) is equivalent to:
\bea
\d_{\mu}t^{\mu}_{a} - m_{a}^{2}~t_{a} = T_{a}
\nonumber \\
t^{\mu}_{a} + \d^{\mu}t_{a} + \d_{\nu}t^{\nu\mu}_{a} = 0
\nonumber \\
t^{\mu\nu}_{a} = t^{\nu\mu}_{a}.
\eea
One can obtain easily from this system that
\be
T_{a} = (\square + m_{a}^{2})~t_{a}.
\ee
Writing a generic form for 
$
t_{a}
$
it is easy to prove that in fact:
\be
T_{a} = 0;
\ee
from here we easily obtain the total antisymmetry of the expressions
$
f^{(1)}_{abc}
$
and
$
f^{(3)}_{abc};
$
also we have
$
g^{(2)}_{abc} = 0.
$
Now one can take 
$
f^{(3)}_{abc} = 0
$
if we subtract from $T$ a total divergence. As a result we obtain the (unique) solution:
\be
T = f^{(1)}_{abc} ( v_{a}^{\mu} v_{b}^{\nu} \partial_{\nu}v_{c\mu}
- v_{a}^{\mu} u_{b} \partial_{\mu}\tilde{u}_{c})
\ee
which is the expression from the theorem.

\section{Perturbation Theory\label{pert}}
Suppose the Wick monomials
$
A_{1},\dots,A_{n}
$
are self-adjoint:
$
A_{j}^{\dagger} = A_{j},~\forall j = 1,\dots,n
$
and of Fermi number
$
f_{i}.
$
We impose the {\it causality} property:
\be
A_{j}(x)~A_{k}(y) = (- 1)^{f_{j}f_{k}}~A_{k}(y)~A_{j}(x)
\ee
for 
$
(x - y)^{2} < 0
$
i.e.
$
x - y
$
outside the causal cones (this relation is denoted by
$
x \sim y
$).

The chronological products
$
T(A_{1}(x_{1}),\dots,A_{n}(x_{n})) \equiv T^{A_{1},\dots,A_{n}}(x_{1},\dots,x_{n}) \quad n = 1,2,\dots
$
are some distribution-valued operators leaving invariant the algebraic Fock space and verifying the following set of axioms:
\begin{itemize}
\item
{\bf Skew-symmetry} in all arguments:
\be
T(\dots,A_{i}(x_{i}),A_{i+1}(x_{i+1}),\dots,) =
(-1)^{f_{i} f_{i+1}} T(\dots,A_{i+1}(x_{i+1}),A_{i}(x_{i}),\dots)
\ee

\item
{\bf Poincar\'e invariance}: we have a natural action of the Poincar\'e group in the
space of Wick monomials and we impose that for all 
$g \in inSL(2,\C)$
we have:
\be
U_{g} T(A_{1}(x_{1}),\dots,A_{n}(x_{n})) U^{-1}_{g} =
T(g\cdot A_{1}(x_{1}),\dots,g\cdot A_{n}(x_{n}))
\label{invariance}
\ee
where in the right hand side we have the natural action of the Poincar\'e group on the set of Wick polynomials.

Sometimes it is possible to supplement this axiom by other invariance
properties: space and/or time inversion, charge conjugation invariance, global
symmetry invariance with respect to some internal symmetry group, supersymmetry,
etc. We will need in the following some of these invariance properties.
\item
{\bf Causality}: if 
$
x - y 
$
is in the upper causal cone then we denote this relation by
$
x \succeq y
$.
Suppose that we have 
$x_{i} \succeq x_{j}, \quad \forall i \leq k, \quad j \geq k+1$;
then we have the factorization property:
\be
T(A_{1}(x_{1}),\dots,A_{n}(x_{n})) =
T(A_{1}(x_{1}),\dots,A_{k}(x_{k}))~~T(A_{k+1}(x_{k+1}),\dots,A_{n}(x_{n}));
\label{causality}
\ee

\item
{\bf Unitarity}: We define the {\it anti-chronological products} using a convenient notation introduced
by Epstein-Glaser, adapted to the Grassmann context. If 
$
X = \{j_{1},\dots,j_{s}\} \subset N \equiv \{1,\dots,n\}
$
is an ordered subset, we define
\be
T(X) \equiv T(A_{j_{1}}(x_{j_{1}}),\dots,A_{j_{s}}(x_{j_{s}})).
\ee
Let us consider some Grassmann variables
$
\theta_{j},
$
of parity
$
f_{j},  j = 1,\dots, n
$
and let us define
\be
\theta_{X} \equiv \theta_{j_{1}} \cdots \theta_{j_{s}}.
\ee
Now let
$
(X_{1},\dots,X_{r})
$
be a partition of
$
N = \{1,\dots,n\}
$
where
$
X_{1},\dots,X_{r}
$
are ordered sets. Then we define the sign
$
\epsilon(X_{1},\dots,X_{r})
$
through the relation
\be
\theta_{1} \cdots \theta_{n} = \epsilon(X_{1}, \dots,X_{r})~\theta_{X_{1}} \dots \theta_{X_{r}}
\ee
and the antichronological products are defined according to
\be
(-1)^{n} \bar{T}(N) \equiv \sum_{r=1}^{n} 
(-1)^{r} \sum_{I_{1},\dots,I_{r} \in Part(N)}
\epsilon(X_{1},\dots,X_{r})~T(X_{1})\dots T(X_{r})
\label{antichrono}
\ee
Then the unitarity axiom is:
\be
\bar{T}(N) = T(N)^{\dagger}.
\label{unitarity}
\ee
\item
{\bf The ``initial condition"}:
\be
T(A(x)) = A(x).
\ee

\item
{\bf Power counting}: We can also include in the induction hypothesis a limitation on the order of
singularity of the vacuum averages of the chronological products associated to
arbitrary Wick monomials
$A_{1},\dots,A_{n}$;
explicitly:
\be
\omega(<\Omega, T^{A_{1},\dots,A_{n}}(X)\Omega>) \leq
\sum_{l=1}^{n} \omega(A_{l}) - 4(n-1)
\label{power}
\ee
where by
$\omega(d)$
we mean the order of singularity of the (numerical) distribution $d$ and by
$\omega(A)$
we mean the canonical dimension of the Wick monomial $W$.
\end{itemize}

Usually, one adds to these set of axioms, the Wick expansions property \cite{EG}.

The basic object of the causal formalism is the causal commutator, defined by:
\be
D(N) \equiv \sum_{(X,Y) \in Part(N)}~( - 1)^{|X|}~\epsilon(X,Y)~[T(Y), \bar{T}(X) ]
\ee
where the partitions
$
(X,Y)
$
are restricted by
$
n \in X, Y \not= \emptyset,
$
$
|Y|
$
is the cardinal of $Y$ and the commutator is graded. These restrictions guarantee that
$
|X|, |Y| < n
$
so the expressions in the right-hand side of the previous expression depend on chronological products in orders
$
\leq n - 1
$
of the perturbation theory. Moreover, it can be proved that the expression
$
D(N) = D(A_{1}(x_{1}),\dots,A_{n}(x_{n}))
$
has causal support in the variables
$
x_{1} - x_{n},\dots,x_{n-1} - x_{n}
$
so by causal splitting, we can obtain the chronological products in order $n$. It is this process of causal splitting which
can produce anomalies.

In second order of the perturbation theory the expression
$
D(N)
$
is indeed the (graded) commutator
\be
D(A_{1}(x),A_{2}(y)) = [ A_{1}(x), A_{2}(y)]
\ee
as we are expecting. The process of obtaining the chronological products is a natural generalization of the process of obtaining Feynman propagators. 
Basically, the (anti)commutator of two quantum fields is (up to some numerical factors) the Pauli-Jordan distribution. This distribution can be spit in two ways.
Due to the property of support in the momentum space it can be split in the positive and negative frequency part:
\be
D_{m}(x) = D_{m}^{(+)}(x) + D_{m}^{(-)}(x).
\ee
Due to the causal support property in the configuration space it can be split in the advanced and retarded parts:
\be
D_{m}(x) = D_{m}^{\rm adv}(x) - D_{m}^{\rm ret}(x).
\ee

Then the Feynman propagator is
\be
D^{F} = D^{\rm ret} + D^{(+)} = D^{\rm adv} - D^{(-)}.
\ee

To describe gauge invariance we introduce the following notation:
\be
\delta T(T^{I_{1}}(x_{1}),\dots,T^{I_{n}}(x_{n})) = 
i \sum_{l=1}^{n} (-1)^{s_{l}} {\partial\over \partial x^{\mu}_{l}}
T(T^{I_{1}}(x_{1}),\dots,T^{I_{l}\mu}(x_{l}),\dots,T^{I_{n}}(x_{n})).
\label{der}
\ee
It is easy to prove that we have:
\be
\delta^{2} = 0
\ee
and
\be
[ d_{Q}, \delta ] = 0.
\ee
Next we define 
\be
s \equiv d_{Q} - i \delta
\ee
such that relation (\ref{gauge-n}) can be rewritten as
\be
sT(T^{I_{1}}(x_{1}),\dots,T^{I_{n}}(x_{n})) = 0.
\label{s-gauge-n}
\ee

We note that if we define
\be
\bar{s} \equiv d_{Q} + i \delta
\ee
we have
\be
s\bar{s} = 0, \qquad \bar{s} s = 0.
\label{s2}
\ee
\newpage 
\section{The Off-Shell Formalism\label{off}}

According to \cite{EG}, Sect. 7, we consider generalized free fields instead of free fields. This means that 
we replace the Pauli-Jordan distribution 
$
D_{m}
$
by some off-shell distribution 
$
D_{m}^{\rm off}
$
which does not verify Klein-Gordon equation but converges in some limit (in the sense of distribution theory) to
$
D_{m}.
$ 
For instance we can take 
\be
D_{m}^{\rm off} \equiv \int d\lambda \rho_{m}(\lambda) D_{\lambda}
\ee
where 
$
\rho_{m}(\lambda)
$
is some function converging in some {\it physical limit} to the distribution 
$
\delta(\lambda - m).
$
In this way all the fields from Section \ref{free} we become {\it generalized free fields} \cite{J} i.e. 
they will verify all properties described there except Klein-Gordon equation.
 
If we keep the definition of the gauge charge unchanged we will loose the property
$
Q^{2} = 0.
$ 
If we keep unchanged the expressions of the interaction Lagrangians from the preceding Section, 
but replace all fields by their off-shell counterparts, we also loose the relations (\ref{descent1}). 
However, these relations will be replaced by 
\be
d_{Q}T^{I} = i~\partial_{\mu}T^{I\mu} + S^{I} \qquad \Leftrightarrow\qquad sT^{I} = S^{I}.
\label{descent1-off}
\ee
with 
$
S^{I}
$
some polynomials which will be null in the on-shell limit. We will need these expressions in the following. 
We will denote
$
K_{c} \equiv K_{m_{c}}
$
and we assume that all fields are off-shell. We have by direct computations the following result:

\begin{thm}
The expressions
$
S^{I} 
$
have the following explicit form:
\be
S = S^{\emptyset} \equiv i~f_{abc}~u_{a}~\left(v_{b}^{\mu}~K_{c}v_{c\mu}
+ {1\over 2}~u_{a}~u_{b}~K_{c}\tilde{u}_{c}\right).
\ee
Also
\be
S^{\mu} \equiv {i\over 2}~f_{abc}~u_{a}~u_{b}~K_{c}v_{c}^{\mu}
\ee
and
\be
S^{I} = 0,\quad |I| > 1. 
\ee
\label{ym-s}
\end{thm}

We now consider the tree contributions to the chronological products. Then we have:
\begin{prop}
In the off-shell formalism we can choose the the second order chronological products such that the following identity is true:
\be
sT^{\rm tree}(T^{I}(x), T^{J}(y)) =  T^{\rm tree}(S^{I}(x),T^{J}(y)) + (-1)^{|I|}~T^{\rm tree}(T^{I}(x),S^{J}(y))
\label{gauge-off}
\ee
\end{prop}
Indeed, we have trivially for the causal commutator:
\be
s D(T^{I}(x), T^{J}(y)) = D(S^{I}(x),T^{J}(y)) + (-1)^{|I|}~D(T^{I}(x),S^{J}(y))
\ee
and if we make the substitution
$
D_{m}^{\rm off} \rightarrow D_{m}^{F,\rm off }
$
we obtain immediately the relation from the statement. 
Now we obtain a clear origin of the anomalies. In the right hand side of (\ref{gauge-off}) there are usually expressions  of the 
type
$
K D_{m}^{F,\rm off}, \partial_{\mu}K D_{m}^{F,\rm off},
$
etc. In the physical limit
we have
\be
K_{m}~D_{m}^{F} = (\square + m^{2})~D^{F}_{m} = \delta(x - y)
\label{kg-df}
\ee
so we obtain in the right hand side of (\ref{gauge-off}) some quasi-local expressions. So we have on-shell:
\be
sT^{\rm tree}(T^{I}(x), T^{J}(y)) = A^{\rm tree}(T^{I}(x), T^{J}(y))
\label{gauge-on}
\ee
where 
$
A^{\rm tree}(T^{I}(x), T^{J}(y))
$
are quasi-local expressions i.e. distribution-valued operators with the support in
$
x = y
$
called {\it anomalies}.
If we can get rid of these anomalies by redefining the chronological products, then the theory is gauge invariant in the second
order of perturbation theory.
We start with the anomaly
$
A^{\rm tree}(T(x), T(y))
$;
by direct computation we can establish that
\be
A^{\rm tree}(T(x), T(y)) = [\delta(x - y) A(x,y) + \partial_{\alpha}\delta(x - y) A^{\alpha}(x,y) ]
+ (x \leftrightarrow y) 
\ee
where:
\bea
A(x,y) = f_{abe} f_{cde} [ - u_{a}(x) v_{b\mu}(x) v_{c\nu}(y) F^{\nu\mu}_{d}(y)
+ u_{a}(x) v_{b}^{\mu}(x) u_{c}(y) \d_{\mu}\tilde{u}_{d}(y)
\nonumber \\
- 1/2~u_{a}(x) u_{b}(x) v_{c}^{\mu}(y) \d_{\mu}\tilde{u}_{d}(y)]
\eea
and
\bea
A^{\alpha}(x,y) = f_{abe} f_{cde} u_{a}(x) v_{b\nu}(x) v_{c}^{\nu}(y) v^{\alpha}_{d}(y).
\eea
By ``integration by parts" we obtain the equivalent form:
\be
A(x,y) = 2~\delta(x - y)~a(x,y) 
+ \left[{\d \over \d x^{\alpha}}a^{\alpha}(x,y) + (x \leftrightarrow y) \right]
\label{ano-b}
\ee
where
\be
a(x,y) \equiv A(x,y) - {\d \over \d x^{\alpha}}A^{\alpha}(x,y),
\qquad
a^{\alpha}(x,y) \equiv \delta(x - y)A^{\alpha}(x,y).
\label{a}
\ee

If we make the redefinition
\be
T(T^{\mu}(x),T(y)) \rightarrow T(T^{\mu}(x),T(y)) + i~a^{\mu}(x,y)
\ee
of the chronological products we will put the anomaly in the form 
\be
A(x,y) = 2~\delta(x - y)~a(x,x) 
\label{ano-c}
\ee

The preceding anomaly can be eliminated iff the expression
$
a(x) = a(x,x) 
$
is a relative cocycle i.e. we have
\be
a = d_{Q}B - i \d_{\mu}B^{\mu}
\label{b}
\ee
for some Wick polynomials $B$ and
$
B^{\mu}
$.
The Wick polynomials
$
B(x)
$
and
$
B^{\mu}(x)
$
are constrained by: (a) Lorentz invariance; (b) ghost number restrictions:
\be
gh(B) = 0,\qquad gh(B^{\mu}) = 1
\ee
and (c) power counting which in our case gives:
\be
\omega(B)~,\omega(B^{\mu}) \leq 4.
\ee
respectively. 

We compute the expression (\ref{a}) and obtain:
\be
a(x,x) = - d_{Q} N + 
( f_{ace}~f_{dbe} + f_{ade}~f_{bce} + f_{abe}~f_{cde} )~
(u_{a}~F_{b\mu\nu}~v_{c}^{\mu}~v_{d}^{\nu} - u_{a}~u_{b}~v_{c}^{\mu}~\partial_{\mu}\tilde{u}_{c} ) 
\label{axx}
\ee
where $N$ is the expression
\be
N = {i\over 2}~f_{abe}~f_{cde}~
v_{a}^{\mu}~v_{b}^{\nu}~v_{c\mu}~v_{d\mu}.
\ee
If we impose the condition (\ref{b}) taking an arbitrary ansatz for $B$ and
$
B^{\mu}
$  
we obtain that the last term in the right hand side of (\ref{axx}) must be null i.e. we have Jacobi identity:
\be
f_{abc}f_{dec} + f_{bdc} f_{aec} + f_{dac} f_{bec} = 0.
\label{Jacobi}
\ee

The renormalized expression of the chronological product
\be
T^{\rm tree(R)}(T(x),T(y)) \equiv T^{\rm tree}(T(x),T(y))+ \delta(x - y)~N(x)
\ee
verifies the second order gauge invariance condition
\be
sT^{\rm tree(R)}(x,y) = 0.
\ee

We can extend the argument for the general second order chronological products: we can have 
gauge invariance condition (\ref{gauge-n}) for 
$
n = 2
$
{\it iff} the constants verify the Jacobi identity. The renormalized the chronological product
\be
T^{\rm tree(R)}(T^{I}(x),T^{J}(y)) \equiv T^{\rm tree}(T^{I}(x),T^{J}(y))+ \delta(x - y)~N^{I,J}(x)
\ee
where
\bea
N^{\emptyset\emptyset} = {i\over 2}~f_{abe}~f_{cde}~
v_{a}^{\mu}~v_{b}^{\nu}~v_{c\mu}~v_{d\mu}
\nonumber\\
N^{[\mu]\emptyset} = - i~f_{abe}~f_{cde}~
u_{a}~v_{b}^{\nu}~v_{c}^{\nu}~v_{d}^{\mu}
\nonumber\\
N^{[\mu][\nu]} = - i~f_{abe}~f_{cde}~
( u_{a}~v_{b}^{\nu}~u_{c}~v_{d}^{\mu} - \eta^{\mu\nu}~u_{a}~v_{b}^{\nu}~u_{c}~v_{d\nu} )
\nonumber\\
N^{[\mu\nu]\emptyset} = - {i\over 2}~f_{abe}~f_{cde}~
u_{a}~u_{b}~v_{c}^{\mu}~v_{d}^{\nu}
\nonumber\\
N^{[\mu\nu][\rho]} = - {i\over 2}~f_{abe}~f_{cde}~
[ \eta^{\mu\rho}~u_{a}~u_{b}~u_{c}~v_{d}^{\nu} - (\mu \leftrightarrow \nu ) ]
\nonumber\\
N^{[\mu\nu][\rho\sigma]} = {i\over 4}~f_{abe}~f_{cde}~
(\eta^{\mu\rho}~\eta^{\nu\sigma} - \eta^{\nu\rho}~\eta^{\mu\sigma})~u_{a}~u_{b}~u_{c}~u_{d}
\eea
verify
\be
sT^{\rm tree(R)}(T^{I}(x),T^{J}(y)) = 0.
\ee
One can prove that gauge invariance is true for loop contributions also \cite{standard}.

\newpage
\section{Anomalies for the $su(2)$ Model\label{ano}}
\begin{thm}
Let us consider a pQFT determined by: 
(i) $g = su(2)$
(ii) parity invariance
(iii) all masses are null.
Then gauge invariance can be imposed in all orders of perturbation theory.
\end{thm}
{\bf Proof:} We can choose the chronological products to be invariant with respect to parity and
$
su(2)
$
in all orders of perturbation theory using the methods presented, for instance in \cite{Sc2}. We prove gauge invariance by induction. 
The first two orders of perturbation theory
have been taken care off in the preceding section. We suppose that we have
\be
sT(T^{I_{1}}(x_{1}),\dots,T^{I_{p}}(x_{p})) = 0, \quad p = 1,\dots,n - 1.
\ee
Then in order $n$ we can have a breakdown of gauge invariance
\be
sT(T^{I_{1}}(x_{1}),\dots,T^{I_{n}}(x_{n})) = A(T^{I_{1}}(x_{1}),\dots,T^{I_{n}}(x_{n}))
\ee
with quasi-local expressions
$
A(T^{I_{1}}(x_{1}),\dots,T^{I_{n}}(x_{n})).
$
These anomalies will be invariant with respect to parity and
$
su(2)
$
and verify some restrictions on the canonical dimension and ghost number:
\be
\omega(A) \leq 5, \quad gh(A(T^{I_{1}}(x_{1}),\dots,T^{I_{n}}(x_{n})) = |I_{1}| + \cdots |I_{n}| + 1.
\label{restrictions}
\ee
Also, if we apply to the preceding equation the operator
$
\bar{s}
$
and take into account that 
\be
\bar{s} s = 0
\ee
we obtain a consistency condition, similar to Wess-Zumino consistency condition from the usual BRST approach.
\be
\bar{s}A(T^{I_{1}}(x_{1}),\dots,T^{I_{n}}(x_{n})) = 0.
\ee

(i) We start a descent procedure. i.e. we first analyze the anomalies of greatest ghost number. From (\ref{restrictions}) we have

\be
A(T^{I_{1}}(x_{1}),\dots,T^{I_{n}}(x_{n})) = 0, \quad |I_{1}| + \cdots |I_{n}| \geq 5.
\ee
So we start with the case
$
\quad |I_{1}| + \cdots |I_{n}| = 4.
$
From (\ref{restrictions}) we have
\be
A(T^{\mu_{1}}(x_{1}),\dots,T^{\mu_{4}}(x_{4}),\dots,T(x_{n})) = \delta(X)~W_{1}^{[\mu_{1},\dots,\mu_{4}]}(x_{1}), \qquad
\quad gh(W_{1}) = 5.
\ee
It follows that we must have:
\be
W_{1}^{[\mu_{1},\dots,\mu_{4}]} = f^{[\mu_{1},\dots,\mu_{4}]}_{a_{1}\dots a_{5}} u_{a_{1}} \dots u_{a_{5}}
\ee
with the numerical Lorentz invariant tensor 
$
f^{[\mu_{1},\dots,\mu_{4}]}_{\dots}
$
completely antisymmetric i.e.
$
f^{[\mu_{1},\dots,\mu_{4}]}_{\dots} \sim \epsilon^{\mu_{1},\dots,\mu_{4}}.
$
But such structure is forbidden by parity invariance so we have:
\be
W_{1} = 0 \quad \Longrightarrow A(T^{\mu_{1}}(x_{1}),\dots,T^{\mu_{4}}(x_{4}),\dots,T(x_{n})) = 0
\ee
Similarly we have
\be
A(T^{\mu_{1}\mu_{2}}(x_{1}),T^{\mu_{3}}(x_{3}),T^{\mu_{4}}(x_{4}),\dots,T(x_{n})) = \delta(X)~W_{2}^{[\mu_{1}\mu_{2}],[\mu_{3}\mu_{4}]}(x_{1}),
\quad gh(W_{2}) = 5
\ee
but now we have the solution
\be
W_{2}^{[\mu_{1}\mu_{2}],[\mu_{3}\mu_{4}]} = ( \eta^{\mu_{1}\mu_{3}}~\eta^{\mu_{2}\mu_{4}} - \eta^{\mu_{1}\mu_{4}}~\eta^{\mu_{2}\mu_{3}} )~
f^{(1)}_{[a_{1}\dots a_{5}]}~u_{a_{1}} \dots u_{a_{5}}.
\ee
Finally we have
\be
A(T^{\mu_{1}\mu_{2}}(x_{1}),T^{\mu_{3}\mu_{4}}(x_{2}),\dots,T(x_{n})) = \delta(X)~W_{3}^{[\mu_{1}\mu_{2}],[\mu_{3}\mu_{4}]}(x_{1}),
\quad gh(W_{2}) = 5
\ee
with the solution
\be
W_{3}^{[\mu_{1}\mu_{2}],[\mu_{3}\mu_{4}]} = ( \eta^{\mu_{1}\mu_{3}}~\eta^{\mu_{2}\mu_{4}} - \eta^{\mu_{1}\mu_{4}}~\eta^{\mu_{2}\mu_{3}} )~
f^{(2)}_{[a_{1}\dots a_{5}]}~u_{a_{1}} \dots u_{a_{5}}.
\ee

(ii) Now we consider the case
$
\quad |I_{1}| + \cdots |I_{n}| = 3.
$
We have the (quasi-local) structure
\bea
A(T^{\mu_{1}}(x_{1}),T^{\mu_{2}}(x_{2}),T^{\mu_{3}}(x_{3}),\dots,T(x_{n}))
\nonumber\\
= \delta(X)~W_{1}^{[\mu_{1}\mu_{2}\mu_{3}]}(x_{1}) 
+ \sum_{j = 1}^{n}~\partial^{j}_{\mu_{4}}\delta(X)~W_{1,j}^{[\mu_{1}\mu_{2}\mu_{3}]\mu_{4}}(X),
\quad gh(W^{\dots}_{1}) = 4.
\label{[123]}
\eea
We can rewrite this anomaly in a much more simpler form: We consider the second term from above as follows
\bea
A_{2} = \sum_{j = 1}^{n}~\{ \partial^{j}_{\mu_{4}}~[\delta(X)~W_{1,j}^{[\mu_{1}\mu_{2}\mu_{3}]\mu_{4}}(X)]
- \delta(X)\partial^{j}_{\mu_{4}}~W_{1,j}^{[\mu_{1}\mu_{2}\mu_{3}]\mu_{4}}(X) \}
\nonumber\\
= \sum_{j = 1}^{n}~ \partial^{j}_{\mu_{4}}~[\delta(X)~W_{1,j}^{[\mu_{1}\mu_{2}\mu_{3}]\mu_{4}}(x_{1})]
- \delta(X)~\sum_{j = 1}^{n}~(\partial^{j}_{\mu_{4}}W_{1,j}^{[\mu_{1}\mu_{2}\mu_{3}]\mu_{4}})(x_{1})
\eea
with
$
W_{1,j}^{\mu_{1}\mu_{2}\mu_{3}\mu_{4}}(x) \equiv W_{1,j}^{\mu_{1}\mu_{2}\mu_{3}\mu_{4}}(x,\dots,x).
$
The last contribution from the previous formula can be eliminated if we redefine the first contribution
$
A_{1}
$
from (\ref{[123]}). We are left with
\be
A_{2} = \sum_{j = 1}^{n}~ [\partial^{j}_{\mu_{4}}~\delta(X)]~W_{1,j}^{[\mu_{1}\mu_{2}\mu_{3}]\mu_{4}}(x_{1})
+ \delta(X)~\partial_{\mu_{4}}~W_{1,1}^{[\mu_{1}\mu_{2}\mu_{3}]\mu_{4}}(x_{1}) 
\ee
Because of the well known identity
\be
{\partial\over \partial x^{\mu}_{j}} \delta(X) = 0
\ee
the first term from above is zero and the last term can be eliminated if we redefine the first contribution
$
A_{1}
$
from (\ref{[123]}); so in the end we can take:
\be
A(T^{\mu_{1}}(x_{1}),T^{\mu_{2}}(x_{2}),T^{\mu_{3}}(x_{3}),\dots,T(x_{n}))
= \delta(X)~W_{1}^{[\mu_{1}\mu_{2}\mu_{3}]}(x_{1}).
\ee
The generic form of the Wick polynomial from the right hand side is
\be
W_{1}^{[\mu_{1}\mu_{2}\mu_{3}]} = a^{[\mu_{1}\mu_{2}\mu_{3}]\mu_{4}}_{[a_{1} a_{2} a_{3}] a_{4}}~
u_{a_{1}} u_{a_{2}} u_{a_{3}} \partial_{\mu_{4}}u_{4}
+ b^{[\mu_{1}\mu_{2}\mu_{3}]\mu_{4}}_{[a_{1} a_{2} a_{3} a_{4}] a_{5}}~u_{a_{1}} u_{a_{2}} u_{a_{3}} u_{a_{4}} v_{a_{5}\mu_{4}}.
\ee
However, because of the antisymmetry properties we have as above
\be
a^{[\mu_{1}\mu_{2}\mu_{3}]\mu_{4}}_{\dots} \sim \epsilon^{\mu_{1},\dots,\mu_{4}}, \quad  
b^{[\mu_{1}\mu_{2}\mu_{3}]\mu_{4}}_{\dots} \sim \epsilon^{\mu_{1},\dots,\mu_{4}} 
\quad \Rightarrow \quad
W_{1}^{[\mu_{1}\mu_{2}\mu_{3}]} = 0 \quad \Rightarrow 
\ee
\be
A(T^{\mu_{1}}(x_{1}),T^{\mu_{2}}(x_{2}),T^{\mu_{3}}(x_{3}),\dots,T(x_{n})) = 0.
\ee
Similarly:
\be
A(T^{\mu_{1}\mu_{2}}(x_{1}),T^{\mu_{3}}(x_{2}),\dots,T(x_{n}))
= \delta(X)~W_{2}^{[\mu_{1}\mu_{2}]\mu_{3}}(x_{1}), \quad gh(W_{2}) = 4
\ee
with
\bea
W_{2}^{[\mu_{1}\mu_{2}]\mu_{3}} = 
f^{(1)}_{[a_{1} a_{2} a_{3}] a_{4}}~\eta^{\mu_{1}\mu_{3}}~u_{a_{1}} u_{a_{2}} u_{a_{3}} \partial^{\mu_{2}}u_{a_{4}} 
- (\mu_{1} \leftrightarrow \mu_{2})
\nonumber\\
+ f^{(2)}_{[a_{1} a_{2} a_{3} a_{4}] a_{5}}~\eta^{\mu_{1}\mu_{3}}~u_{a_{1}} u_{a_{2}} u_{a_{3}} u_{a_{4}} v_{a_{5}}^{\mu_{2}}
- (\mu_{1} \leftrightarrow \mu_{2}).
\eea
Now we impose the Wess-Zumino consistency condition:
\be
\bar{s}A(T^{\mu_{1}\mu_{2}}(x_{1}),T^{\mu_{3}}(x_{2}),\dots) = 0
\ee
or, in detail:
\bea
d_{Q}A(T^{\mu_{1}\mu_{2}}(x_{1}),T^{\mu_{3}}(x_{2}),\dots)
+ i~{\partial\over \partial x^{\mu_{4}}_{2}}A(T^{\mu_{1}\mu_{2}}(x_{1}),T^{\mu_{3}\mu_{4}}(x_{2}),\dots)
\nonumber\\
- i \sum_{l=3}^{n}~{\partial\over \partial x^{\mu_{4}}_{l}}A(T^{\mu_{1}\mu_{2}}(x_{1}),T^{\mu_{3}}(x_{2}),\dots,T^{\mu_{4}}(x_{l}),\dots)  = 0.
\eea
A simple computation gives the following relations:
\bea
f^{(1)}_{a_{1} a_{2} a_{3} a_{4}} = 0
\nonumber\\
f^{(1)}_{a_{1} a_{2} a_{3} a_{4}} + f^{(2)}_{a_{1} a_{2} a_{3} a_{4}} = 0
\nonumber\\
d_{Q}W_{2}^{[\mu_{1}\mu_{2}]\mu_{3}} = 0 .
\eea
From the first two relations we have
\be
f^{(j)}_{a_{1},\dots,a_{4}} = 0, \quad j = 1 ,2
\ee
and from the last:
\be
f^{(2)}_{a_{1} a_{2} a_{3} a_{4} a_{5}} = 0.
\ee
It follows that
\be
A(T^{I_{1}}(x_{1}),\dots,T^{I_{n}}(x_{n})) = 0, \quad |I_{1}| + \cdots |I_{n}| = 4
\ee
and
\bea
W_{2}^{[\mu_{1}\mu_{2}]\mu_{3}} = f^{(1)}_{[a_{1} a_{2} a_{3}] a_{4}}~\eta^{\mu_{1}\mu_{3}}~u_{a_{1}} u_{a_{2}} u_{3} \partial^{\mu_{2}}u_{4} 
- (\mu_{1} \leftrightarrow \mu_{2})
\nonumber\\
= i~f^{(1)}_{[a_{1} a_{2} a_{3}] a_{4}}~\eta^{\mu_{1}\mu_{3}}~d_{Q} [u_{a_{1}} u_{a_{2}} u_{3} v^{\mu_{2}}_{a_{4}} ]
- (\mu_{1} \leftrightarrow \mu_{2}).
\eea
So, if we redefine conveniently the expression
$
T(T^{\mu_{1}\mu_{2}}(x_{1}),T^{\mu_{3}}(x_{2}),\dots,T(x_{n}))
$
we can fix:
\be
A(T^{I_{1}}(x_{1}),\dots,T^{I_{n}}(x_{n})) = 0, \quad |I_{1}| + \cdots |I_{n}| = 3.
\ee

(iii) Next we consider the case 
$
\quad |I_{1}| + \cdots |I_{n}| = 2.
$
We have the (quasi-local) structure
\bea
A(T^{\mu_{1}}(x_{1}),T^{\mu_{2}}(x_{2}),\dots,T(x_{n}))
\nonumber\\
= \delta(X)~W_{1}^{[\mu_{1}\mu_{2}]}(x_{1}) 
+ \sum_{j = 1}^{n}~\partial^{j}_{\mu_{3}}\delta(X)~W_{1,j}^{[\mu_{1}\mu_{2}]\mu_{3}}(X)
\nonumber\\
+ \sum_{j,k = 1}^{n}~\partial^{j}_{\mu_{3}}\partial^{k}_{\mu_{4}}\delta(X)~W_{1,jk}^{[\mu_{1}\mu_{2}]\mu_{3}\mu_{4}}(X),
\quad gh(W^{\dots}_{1}) = 3
\eea
As at (ii) we can rewrite it in the form:
\be
A(T^{\mu_{1}}(x_{1}),T^{\mu_{2}}(x_{2}),\dots,T(x_{n})) = \delta(X)~W_{1}^{[\mu_{1}\mu_{2}]}(x_{1}).
\ee
The generic form of the Wick polynomial from the right hand side is
\bea
W_{1}^{[\mu_{1}\mu_{2}]} = L_{1}^{[\mu_{1}\mu_{2}]}~f^{(1)}_{[a_{1}a_{2}a_{3}]}~u_{a_{1}} u_{a_{2}} u_{a_{3}} 
+ L_{2}^{[\mu_{1}\mu_{2}]\mu_{3}}~f^{(2)}_{[a_{1}a_{2}]a_{3}}~u_{a_{1}} u_{a_{2}} \partial_{\mu_{3}}u_{a_{3}} 
\nonumber\\
+ L_{3}^{[\mu_{1}\mu_{2}]\mu_{3}\mu_{4}}~f^{(3)}_{a_{1}a_{2}a_{3}}~u_{a_{1}} \partial_{\mu_{3}}u_{a_{2}} \partial_{\mu_{4}}u_{a_{3}} 
+ L_{4}^{[\mu_{1}\mu_{2}]\mu_{3}\mu_{4}}~f^{(4)}_{[a_{1}a_{2}]a_{3}}~u_{a_{1}} u_{a_{2}} \partial_{\mu_{3}}\partial_{\mu_{4}}u_{a_{3}} 
\nonumber\\
+ L_{5}^{[\mu_{1}\mu_{2}]\mu_{3}}~f^{(5)}_{[a_{1}a_{2}a_{3}]a_{4}}~u_{a_{1}} u_{a_{2}} u_{a_{3}} v_{a_{4}\mu_{3}}
+ L_{6}^{[\mu_{1}\mu_{2}]\mu_{3}\mu_{4}}~f^{(6)}_{[a_{1}a_{2}]a_{3}a_{4}}~u_{a_{1}} u_{a_{2}} \partial_{\mu_{3}}u_{a_{3}} v_{a_{4}\mu_{4}}
\nonumber\\
+ L_{7}^{[\mu_{1}\mu_{2}]\mu_{3}\mu_{4}}~f^{(7)}_{[a_{1}a_{2}a_{3}]a_{4}}~u_{a_{1}} u_{a_{2}} u_{a_{3}} \partial_{\mu_{3}}v_{a_{4}\mu_{4}}
+ L_{8}^{[\mu_{1}\mu_{2}]\mu_{3}\mu_{4}}~f^{(8)}_{[a_{1}a_{2}a_{3}]a_{4}a_{5}}~u_{a_{1}} u_{a_{2}} u_{a_{3}} v_{a_{4}\mu_{3}} v_{a_{5}\mu_{4}}
\nonumber\\
+ L_{9}^{[\mu_{1}\mu_{2}]}~f^{(9)}_{[a_{1}a_{2}a_{3}a_{4}]a_{5}}~u_{a_{1}} u_{a_{2}} u_{a_{3}} u_{a_{4}} \tilde{u}_{a_{5}}
\eea
Lorentz covariance considerations give us
\be
L^{\dots}_{j} = 0, \quad j = 2, 5
\ee
and antisymmetry considerations lead to
\be
L^{\dots}_{j} = 0, \quad j = 1, 9
\ee
and
\be
L_{j}^{[\mu_{1}\mu_{2}]\mu_{3}\mu_{4}} \sim ( \eta^{\mu_{1}\mu_{3}}~\eta^{\mu_{2}\mu_{4}} - \eta^{\mu_{1}\mu_{4}}~\eta^{\mu_{2}\mu_{3}} ),
\quad j = 3, 4, 6, 7, 8
\ee
so
$
L^{\dots}_{4} = 0.
$
In the end we have (after redefining the numerical tensors)
\bea
W_{1}^{[\mu_{1}\mu_{2}]} = f^{(1)}_{a_{1}a_{2}a_{3}}~u_{a_{1}} \partial^{\mu_{1}}u_{a_{2}} \partial^{\mu_{2}}u_{a_{3}} 
- (\mu_{1} \leftrightarrow \mu_{2})
\nonumber\\
+ f^{(2)}_{[a_{1}a_{2}]a_{3}a_{4}}~u_{a_{1}} u_{a_{2}} \partial^{\mu_{1}}u_{a_{3}} v_{a_{4}}^{\mu_{2}}
- (\mu_{1} \leftrightarrow \mu_{2})
\nonumber\\
+ f^{(3)}_{[a_{1}a_{2}a_{3}]a_{4}} u_{a_{1}} u_{a_{2}} u_{a_{3}} F_{a_{4}}^{\mu_{1}\mu_{2}}
+ f^{(4)}_{[a_{1}a_{2}a_{3}][a_{4}a_{5}]} u_{a_{1}} u_{a_{2}} u_{a_{3}} v_{a_{4}}^{\mu_{1}} v_{a_{5}}^{\mu_{2}}
\eea
In a similar way we argue that we have
\be
A(T^{\mu_{1}\mu_{2}}(x_{1}),\dots,T(x_{n})) = \delta(X)~W_{2}^{[\mu_{1}\mu_{2}]}(x_{1}) 
\ee
with
\bea
W_{2}^{[\mu_{1}\mu_{2}]} = g^{(1)}_{a_{1}a_{2}a_{3}}~u_{a_{1}} \partial^{\mu_{1}}u_{a_{2}} \partial^{\mu_{2}}u_{a_{3}} 
- (\mu_{1} \leftrightarrow \mu_{2})
\nonumber\\
+ g^{(2)}_{[a_{1}a_{2}]a_{3}]a_{4}}~u_{a_{1}} u_{a_{2}} \partial^{\mu_{1}}u_{a_{3}} v_{a_{4}}^{\mu_{2}}
- (\mu_{1} \leftrightarrow \mu_{2})
\nonumber\\
+ g^{(3)}_{[a_{1}a_{2}a_{3}]a_{4}} u_{a_{1}} u_{a_{2}} u_{a_{3}} F_{a_{4}}^{\mu_{1}\mu_{2}}
+ g^{(4)}_{[a_{1}a_{2}a_{3}][a_{4}a_{5}]} u_{a_{1}} u_{a_{2}} u_{a_{3}} v_{a_{4}}^{\mu_{1}} v_{a_{5}}^{\mu_{2}}
\eea
Now we impose the Wess-Zumino consistency relation:
\be
\bar{s}A(T^{\mu_{1}}(x_{1}),T^{\mu_{2}}(x_{2}),\dots) = 0
\ee
which reduces to
\be
d_{Q}A(T^{\mu_{1}}(x_{1}),T^{\mu_{2}}(x_{2}),\dots) = 0
\ee
because of the previous results. From here we have
\be
d_{Q}W_{1}^{[\mu_{1}\mu_{2}]} = 0
\ee
which gives us:
\be 
f^{(2)}_{a_{1}a_{2}a_{3}a_{4}} = - (\mu_{3} \leftrightarrow \mu_{4}), \qquad
f^{(4)}_{a_{1}a_{2}a_{3}a_{4}a_{5}} = 0.
\ee
We are left with
\bea
W_{1}^{[\mu_{1}\mu_{2}]} = i d_{Q} \{ [ f^{(1)}_{a_{1}a_{2}a_{3}}~u_{a_{1}} v_{a_{2}}^{\mu_{1}} \partial^{\mu_{2}}u_{a_{3}} 
- (\mu_{1} \leftrightarrow \mu_{2}) ]
- f^{(2)}_{[a_{1}a_{2}][a_{3}a_{4}]}~u_{a_{1}} u_{a_{2}} v_{a_{3}}^{\mu_{1}} v_{a_{4}}^{\mu_{2}} \}
\nonumber\\
+ f^{(3)}_{[a_{1}a_{2}a_{3}]a_{4}} u_{a_{1}} u_{a_{2}} u_{a_{3}} F_{a_{4}}^{\mu_{1}\mu_{2}}
\eea
so if we redefine the chronological products
$
T(T^{\mu_{1}}(x_{1}), T^{\mu_{2}}(x_{2}), \dots,T(x_{n}))
$
we can take:
\be
W_{1}^{[\mu_{1}\mu_{2}]} = f^{(3)}_{[a_{1}a_{2}a_{3}]a_{4}} u_{a_{1}} u_{a_{2}} u_{a_{3}} F_{a_{4}}^{\mu_{1}\mu_{2}}
\ee
In a similar way, if we redefine the chronological products
$
T(T^{\mu_{1}\mu_{2}}(x_{1}), \dots,T(x_{n}))
$
we can arrange such that
\be
W_{2}^{[\mu_{1}\mu_{2}]} = g^{(3)}_{[a_{1}a_{2}a_{3}]a_{4}} u_{a_{1}} u_{a_{2}} u_{a_{3}} F_{a_{4}}^{\mu_{1}\mu_{2}}
\ee

(iv) In the case 
$
\quad |I_{1}| + \cdots |I_{n}| = 1
$
we have only 
\be
A(T^{\mu_{1}}(x_{1}),\dots,T(x_{n})) = \delta(X)~W^{\mu_{1}}(x_{1}) + \cdots  \qquad gh(W^{\mu}) = 2
\ee
where by 
$\dots$ 
we mean terms with derivatives on the delta distribution. As before, we can skip these terms, so we take:
\be
A(T^{\mu_{1}}(x_{1}),\dots,T(x_{n})) = \delta(X)~W^{\mu_{1}}(x_{1}).
\ee
The generic form of the Wick polynomial from the right hand side is:
\bea
W^{\mu_{1}} = L_{1}^{\mu_{1}\mu_{2}}~f^{(1)}_{a_{1}a_{2}}~u_{a_{1}}  \partial_{\mu_{2}}u_{a_{2}}
+ L_{2}^{\mu_{1},\dots,\mu_{4}}~f^{(2)}_{a_{1}a_{2}} u_{a_{1}} \partial_{\mu_{2}}\partial_{\mu_{3}}\partial_{\mu_{4}}u_{a_{2}} 
\nonumber\\
+ L_{3}^{\mu_{1},\dots,\mu_{4}}~f^{(3)}_{a_{1}a_{2}}~\partial_{\mu_{2}}u_{a_{1}} \partial_{\mu_{3}}\partial_{\mu_{4}}u_{a_{2}} 
+ L_{4}^{\mu_{1}\mu_{2}}~f^{(4)}_{[a_{1}a_{2}]a_{3}}~u_{a_{1}} u_{a_{2}} v_{a_{3}\mu_{2}} 
\nonumber\\
+ L_{5}^{\mu_{1},\dots,\mu_{4}}~f^{(5)}_{a_{1}a_{2}a_{3}}~u_{a_{1}} \partial_{\mu_{2}}\partial_{\mu_{3}}u_{a_{2}} v_{a_{3}\mu_{4}}
+ L_{6}^{\mu_{1},\dots,\mu_{4}}~f^{(6)}_{a_{1}a_{2}a_{3}}~u_{a_{1}} \partial_{\mu_{2}}u_{a_{2}} \partial_{\mu_{3}}v_{a_{3}\mu_{4}}
\nonumber\\
+ L_{7}^{\mu_{1}\mu_{2}}~f^{(7)}_{[a_{1}a_{2}]a_{3}}~u_{a_{1}} u_{a_{2}}  \partial_{\mu_{2}}\partial_{\mu_{3}}v_{a_{3}\mu_{4}}
+ L_{8}^{\mu_{1},\dots,\mu_{4}}~f^{(8)}_{a_{1}a_{2}a_{3}a_{4}}~u_{a_{1}} \partial_{\mu_{2}}u_{a_{2}} v_{a_{3}\mu_{3}} v_{a_{4}\mu_{4}}
\nonumber\\
+ L_{9}^{\mu_{1},\dots,\mu_{4}}~f^{(9)}_{[a_{1}a_{2}]a_{3}a_{4}}~u_{a_{1}} u_{a_{2}} v_{a_{3}\mu_{2}} \partial_{\mu_{3}}v_{a_{4}\mu_{4}}
+ L_{10}^{\mu_{1}\mu_{2}}~f^{(10)}_{[a_{1}a_{2}]a_{3}a_{4}}~u_{a_{1}} u_{a_{2}} \partial_{\mu_{2}}u_{a_{3}} \tilde{u}_{a_{4}}
\nonumber\\
+ L_{11}^{\mu_{1}\mu_{2}}~f^{(11)}_{[a_{1}a_{2}a_{3}]a_{4}}~u_{a_{1}} u_{a_{2}} u_{a_{3}} \partial_{\mu_{2}}\tilde{u}_{a_{4}}
+ L_{12}^{\mu_{1},\dots,\mu_{4}}~f^{(12)}_{[a_{1}a_{2}]a_{3}a_{4}a_{5}}~u_{a_{1}} u_{a_{2}} v_{a_{3}\mu_{2}} v_{a_{4}\mu_{3}} v_{a_{5}\mu_{4}}
\eea
but Lorentz covariance considerations lead us to a more precise form:
\bea
W^{\mu} = f^{(1)}_{a_{1}a_{2}}~u_{a_{1}}  \partial^{\mu}u_{a_{2}}
+ f^{(2)}_{a_{1}a_{2}} \partial_{\nu}u_{a_{1}} \partial^{\mu}\partial^{\nu}u_{a_{2}} 
\nonumber\\
+ f^{(3)}_{[a_{1}a_{2}]a_{3}}~u_{a_{1}} u_{a_{2}} v_{a_{3}}^{\mu}
+ f^{(4)}_{a_{1}a_{2}a_{3}}~u_{a_{1}} \partial^{\mu}\partial_{\nu}u_{a_{2}} v_{a_{3}}^{\nu} 
\nonumber\\
+ f^{(5)}_{a_{1}a_{2}a_{3}}~u_{a_{1}} \partial^{\mu}u_{a_{2}} \partial_{\nu}v_{a_{3}}^{\nu}
+ f^{(6)}_{a_{1}a_{2}a_{3}}~u_{a_{1}} \partial_{\nu}u_{a_{2}} \partial^{\mu}v_{a_{3}}^{\nu}
\nonumber\\
+ f^{(7)}_{a_{1}a_{2}a_{3}}~u_{a_{1}} \partial_{\nu}u_{a_{2}} \partial^{\nu}v_{a_{3}}^{\mu}
+ f^{(8)}_{[a_{1}a_{2}]a_{3}}~u_{a_{1}} u_{a_{2}} \partial^{\mu}\partial_{\nu}v_{a_{3}}^{\nu}
\nonumber\\
+ f^{(9)}_{a_{1}a_{2}\{a_{3}a_{4}\}}~u_{a_{1}} \partial^{\mu}u_{a_{2}} v_{a_{3}}^{\nu} v_{a_{4}\nu}
+ f^{(10)}_{a_{1}a_{2}a_{3}a_{4}}~u_{a_{1}} \partial_{\nu}u_{a_{2}} v_{a_{3}}^{\mu} v_{a_{4}}^{\nu}
\nonumber\\
+ f^{(11)}_{[a_{1}a_{2}]a_{3}a_{4}}~u_{a_{1}} u_{a_{2}} v_{a_{3}}^{\mu} \partial_{\nu}v_{a_{4}}^{\nu}
+ f^{(12)}_{[a_{1}a_{2}]a_{3}a_{4}}~u_{a_{1}} u_{a_{2}} v_{a_{3}}^{\nu} \partial_{\nu}v_{a_{4}}^{\mu}
\nonumber\\
+ f^{(13)}_{[a_{1}a_{2}]a_{3}a_{4}}~u_{a_{1}} u_{a_{2}} v_{a_{3}\nu} \partial^{\mu}v_{a_{4}}^{\nu}
+ f^{(14)}_{[a_{1}a_{2}]a_{3}a_{4}}~u_{a_{1}} u_{a_{2}} \partial^{\mu}u_{a_{3}} \tilde{u}_{a_{4}}
\nonumber\\
+ f^{(15)}_{[a_{1}a_{2}a_{3}]a_{4}}~u_{a_{1}} u_{a_{2}} u_{a_{3}} \partial^{\mu}\tilde{u}_{a_{4}}
+ f^{(16)}_{[a_{1}a_{2}]a_{3}\{a_{4}a_{5}\}}~u_{a_{1}} u_{a_{2}} v_{a_{3}}^{\mu} v_{a_{4}}^{\nu} v_{a_{5}\nu}
\eea
and we now impose the Wess-Zumino consistency condition:
\be
\bar{s}A(T^{\mu_{1}}(x_{1}),\dots) = 0
\ee
or, in detail:
\bea
d_{Q}A(T^{\mu_{1}}(x_{1}),\dots)
\nonumber\\
+ i~{\partial\over \partial x^{\mu_{2}}_{2}}A(T^{\mu_{1}\mu_{2}}(x_{1}),\dots)
- i \sum_{l=2}^{n}~{\partial\over \partial x^{\mu_{2}}_{l}}A(T^{\mu_{1}}(x_{1}),\dots,T^{\mu_{2}}(x_{l}),\dots)  = 0.
\eea
This equation easily leads to
\bea
d_{Q}W^{\mu_{1}} + i~\partial_{\mu_{2}}W_{1}^{\mu_{1}\mu_{2}} = 0
\nonumber\\
W_{2}^{\mu_{1}\mu_{2}} = - W_{1}^{\mu_{1}\mu_{2}}.
\eea
From the second one we obtain:
\be
g^{(3)}_{a_{1} \dots a_{4}} = - f^{(3)}_{a_{1} \dots a_{4}}
\ee
so, for simplicity we denote:
\be
g_{a_{1}\dots a_{4}} \equiv f^{(3)}_{a_{1}\dots a_{4}}
\ee
The first equation leads us to the following system:
\bea
f^{(3)}_{a_{1}a_{2}a_{3}} = 0  
\nonumber\\
f^{(4)}_{a_{1}a_{2}a_{3}} - f^{(6)}_{a_{1}a_{3}a_{2}} -f^{(7)}_{a_{1}a_{3}a_{2}} = 0
\nonumber\\
2 f^{(9)}_{a_{1}a_{2}a_{3}a_{4}} - f^{(10)}_{a_{1}a_{3}a_{2}a_{4}} = 0
\nonumber\\
f^{(10)}_{a_{1}a_{2}a_{3}a_{4}} = a_{2} \leftrightarrow a_{4}
\nonumber\\
f^{(11)}_{a_{1}a_{2}a_{3}a_{4}} + f^{(14)}_{a_{1}a_{2}a_{3}a_{4}} + g_{a_{1}a_{2}a_{3}a_{4}} = 0
\nonumber\\
f^{(12)}_{a_{1}a_{2}a_{3}a_{4}} - g_{a_{1}a_{2}a_{3}a_{4}} = 0
\nonumber\\
f^{(12)}_{a_{1}a_{2}a_{4}a_{3}} + f^{(13)}_{a_{1}a_{2}a_{4}a_{3}} = 0
\nonumber\\
f^{(13)}_{a_{1}a_{2}a_{3}a_{4}} = 0
\nonumber\\
f^{(15)}_{a_{1}a_{2}a_{3}a_{4}} + g_{a_{1}a_{2}a_{3}a_{4}} = 0
\nonumber\\
f^{(16)}_{a_{1}a_{2}a_{3}a_{4}a_{5}} = 0
\nonumber\\
f^{(16)}_{a_{1}a_{2}a_{4}a_{5}a_{3}} = 0.
\eea
where we have computed the expression
$
d_{Q}W^{\mu_{1}} + i~\partial_{\mu_{2}}W_{1}^{\mu_{1}\mu_{2}}
$
and obtained the following (linear independent) Wick monomials:
\bea
u_{a_{1}} u_{a_{2}} \partial^{\mu}u_{a_{3}}, u_{a_{1}} \partial^{\mu}\partial^{\nu}u_{a_{2}} \partial_{\nu}u_{a_{3}},
u_{a_{1}} \partial^{\mu}u_{a_{2}} \partial_{\nu}u_{a_{3}} v_{a_{4}}^{\nu}, 
u_{a_{1}} \partial^{\nu}u_{a_{2}} \partial_{\nu}u_{a_{3}} v_{a_{4}}^{\mu},
u_{a_{1}} u_{a_{2}} \partial^{\mu}u_{a_{3}} \partial^{\nu}v_{a_{4}}^{\nu},
\nonumber\\
u_{a_{1}} u_{a_{2}} \partial_{\nu}u_{a_{3}} \partial^{\nu}v_{a_{4}}^{\mu},
u_{a_{1}} u_{a_{2}} \partial^{\mu}\partial_{\nu}u_{a_{3}} v_{a_{4}}^{\nu},
u_{a_{1}} u_{a_{2}} \partial_{\nu}u_{a_{3}} \partial^{\mu}v_{a_{4}}^{\nu},
u_{a_{1}} u_{a_{2}} u_{a_{3}} \partial^{\mu}\partial_{\nu}v_{a_{4}}^{\nu},
\nonumber\\
u_{a_{1}} u_{a_{2}} \partial^{\mu}u_{a_{3}} v_{a_{4}}^{\nu} v_{a_{5}\nu},
u_{a_{1}} u_{a_{2}} \partial_{\nu}u_{a_{3}} v_{a_{4}}^{\mu} v_{a_{5}}^{\nu}.
\nonumber
\eea
The equations of the system are just the numerical constants of these Wick monomials. The solution of this system is:
\bea
f^{(j)}_{\dots} = 0, \quad j = 3, 12, 13, 16
\nonumber\\
g_{a_{1}a_{2}a_{3}a_{4}} = 0
\nonumber\\
f^{(4)}_{a_{1}a_{2}a_{3}} = f^{(6)}_{a_{1}a_{3}a_{2}} -f^{(7)}_{a_{1}a_{3}a_{2}}
\nonumber\\
f^{(10)}_{a_{1}a_{2}a_{3}a_{4}} = 2 f^{(9)}_{a_{1}a_{3}a_{2}a_{4}} 
\nonumber\\
f^{(14)}_{a_{1}a_{2}a_{3}a_{4}} = - f^{(11)}_{a_{1}a_{2}a_{3}a_{4}}
\nonumber\\
f^{(9)}_{a_{1}a_{2}a_{3}a_{4}} = a_{3} \leftrightarrow a_{4}
\eea
From
$
g_{a_{1}a_{2}a_{3}a_{4}} = 0
$
it follows that 
\be
A(T^{I_{1}}(x_{1}),\dots,T^{I_{n}}(x_{n})) = 0, \quad |I_{1}| + \cdots |I_{n}| = 2.
\ee

The rest of the equations can be used to exhibit the Wick polynomial from the expression of the anomaly in a simpler form:
\bea
W^{\mu} = f^{(1)}_{a_{1}a_{2}}~u_{a_{1}}  \partial^{\mu}u_{a_{2}}
+ f^{(2)}_{a_{1}a_{2}} \partial_{\nu}u_{a_{1}} \partial^{\mu}\partial^{\nu}u_{a_{2}} 
\nonumber\\
+ f^{(6)}_{a_{1}a_{2}a_{3}}~(u_{a_{1}} \partial_{\nu}u_{a_{2}} \partial^{\mu}v_{a_{3}}^{\nu}
+ u_{a_{1}} v_{a_{2}}^{\nu} \partial^{\mu}\partial_{\nu}u_{a_{3}} )
\nonumber\\
+ f^{(7)}_{a_{1}a_{2}a_{3}}~( u_{a_{1}} \partial_{\nu}u_{a_{2}} \partial^{\nu}v_{a_{3}}^{\mu}
+ u_{a_{1}} v_{a_{2}}^{\nu} \partial^{\mu}\partial_{\nu}u_{a_{3}} )
\nonumber\\
+ f^{(8)}_{a_{1}a_{2}a_{3}}~u_{a_{1}} u_{a_{2}} \partial^{\mu}\partial_{\nu}v_{a_{3}}^{\nu}
\nonumber\\
+ f^{(9)}_{a_{1}a_{2}a_{3}a_{4}}~( u_{a_{1}} \partial^{\mu}u_{a_{2}} v_{a_{3}}^{\nu} v_{a_{4}\nu}
+ 2 u_{a_{1}} v_{a_{2}}^{\mu} \partial_{\nu}u_{a_{3}} v_{a_{4}}^{\nu} )
\nonumber\\
+ f^{(11)}_{a_{1}a_{2}a_{3}a_{4}}~(u_{a_{1}} u_{a_{2}} v_{a_{3}}^{\mu} \partial_{\nu}v_{a_{4}}^{\nu}
- u_{a_{1}} u_{a_{2}} \partial^{\mu}u_{a_{3}} \tilde{u}_{a_{4}} ).
\eea
This expression is in fact a coboundary:
\bea
W^{\mu} = i~d_{Q} (f^{(1)}_{a_{1}a_{2}}~u_{a_{1}} v_{a_{2}}^{\mu}
+ f^{(2)}_{a_{1}a_{2}} \partial_{\nu}u_{a_{1}} \partial^{\mu}v_{a_{2}}^{\nu} 
+ f^{(6)}_{a_{1}a_{2}a_{3}}~u_{a_{1}} v_{a_{2}\nu} \partial^{\mu}v_{a_{3}}^{\nu}
+ f^{(7)}_{a_{1}a_{2}a_{3}}~u_{a_{1}} v_{a_{2}}^{\nu} \partial^{\nu}v_{a_{3}}^{\mu}
\nonumber\\
+ f^{(8)}_{a_{1}a_{2}a_{3}}~u_{a_{1}} u_{a_{2}} \partial^{\mu}\tilde{u}_{a_{3}}
+ f^{(9)}_{a_{1}a_{2}a_{3}a_{4}}~u_{a_{1}} v_{a_{2}}^{\mu} v_{a_{3}}^{\nu} v_{a_{4}\nu}
- f^{(11)}_{a_{1}a_{2}a_{3}a_{4}}~u_{a_{1}} u_{a_{2}} v_{a_{3}}^{\mu} \tilde{u}_{a_{4}} )
\eea
so if we redefine the chronological products
$
T(T^{\mu}(x_{1}), \dots)
$
we can also fix:
\be
A(T^{\mu_{1}}(x_{1}),\dots) = 0. 
\ee

(v) Finally we consider the anomaly
\be
A(T(x_{1}),\dots,T(x_{n})) = \delta(X)~W(x_{1}) + \cdots \qquad gh(W^{\mu}) = 1
\ee
where by 
$\dots$ 
we mean terms with derivatives on the delta distribution. As before, we can skip these terms, so we take:
\be
A(T(x_{1}),\dots,T(x_{n})) = \delta(X)~W(x_{1}).
\ee
The generic form of the Wick polynomial from the right hand side is:
\bea
W = L_{1}^{\mu\nu}~f^{(1)}_{a_{1}a_{2}}~\partial_{\mu}u_{a_{1}} v_{a_{2}\nu}
+ L_{2}^{\mu\nu}~f^{(2)}_{a_{1}a_{2}} u_{a_{1}} \partial_{\mu}v_{a_{2}\nu}
\nonumber\\
+ L_{3}^{\mu\nu}~f^{(3)}_{a_{1}a_{2}a_{3}}~u_{a_{1}} v_{a_{2}\mu} v_{a_{3}\nu}
+ L_{4}^{\mu\nu\rho\sigma}~f^{(4)}_{a_{1}a_{2}a_{3}}~\partial_{\mu}\partial_{\nu}u_{a_{1}} v_{a_{2}\rho} v_{a_{3}\sigma} 
\nonumber\\
+ L_{5}^{\mu\nu\rho\sigma}~f^{(5)}_{a_{1}a_{2}a_{3}}~\partial_{\mu}u_{a_{1}} v_{a_{2}\nu} \partial_{\rho}v_{a_{3}\sigma}
+ L_{6}^{\mu\nu\rho\sigma}~f^{(6)}_{a_{1}a_{2}a_{3}}~u_{a_{1}} v_{a_{2}\mu} \partial_{\nu}\partial_{\rho}v_{a_{3}\sigma}
\nonumber\\
+ L_{7}^{\mu\nu\rho\sigma}~f^{(7)}_{a_{1}a_{2}a_{3}}~u_{a_{1}} \partial_{\mu}v_{a_{2}\nu} \partial_{\rho}v_{a_{3}\sigma}
+ f^{(8)}_{a_{1}a_{2}a_{3}}~u_{a_{1}} u_{a_{2}} \tilde{u}_{a_{3}}
\nonumber\\
+ L_{9}^{\mu,\nu}~f^{(9)}_{a_{1}a_{2}a_{3}}~u_{a_{1}} \partial_{\mu}\partial_{\nu}u_{a_{2}} \tilde{u}_{a_{3}}
+ L_{10}^{\mu\nu}~f^{(10)}_{a_{1}a_{2}a_{3}}~u_{a_{1}} u_{a_{2}} \partial_{\mu}\partial_{\nu}\tilde{u}_{a_{3}}
\nonumber\\
+ L_{11}^{\mu\nu}~f^{(11)}_{a_{1}a_{2}a_{3}}~u_{a_{1}} \partial_{\mu}u_{a_{2}} \partial_{\nu}\tilde{u}_{a_{3}}
+ L_{12}^{\mu\nu\rho\sigma}~f^{(12)}_{a_{1}a_{2}a_{3}a_{4}}~\partial_{\mu}u_{a_{1}} v_{a_{2}\nu} v_{a_{3}\rho} v_{a_{4}\sigma}
\nonumber\\
+ L_{13}^{\mu\nu\rho\sigma}~f^{(13)}_{a_{1}a_{2}a_{3}a_{4}}~u_{a_{1}} v_{a_{2}\mu} v_{a_{3}\nu} \partial_{\rho}v_{a_{4}\sigma}
+ L_{14}^{\mu\nu\rho\sigma}~f^{(14)}_{a_{1}a_{2}a_{3}a_{4}a_{5}}~u_{a_{1}} v_{a_{2}\mu} v_{a_{3}\nu} v_{a_{4}\rho} v_{a_{5}\sigma}
\nonumber\\
+ f^{(15)}_{a_{1}a_{2}a_{3}a_{4}a_{5}}~u_{a_{1}} u_{a_{2}} u_{a_{3}} \tilde{u}_{a_{4}} \tilde{u}_{a_{5}}.
\eea
If we add a coboundary and redefine $L_{5}$ we can make
$
L_{4} = 0.
$
Lorentz covariance leads to the following precise form:
\bea
W = f^{(1)}_{a_{1}a_{2}}~\partial_{\mu}u_{a_{1}} v_{a_{2}}^{\mu}
+ f^{(2)}_{a_{1}a_{2}} u_{a_{1}} \partial_{\mu}v_{a_{2}}^{\mu}
\nonumber\\
+ f^{(3)}_{a_{1}\{a_{2}a_{3}\}}~u_{a_{1}} v_{a_{2}\mu} v_{a_{3}}^{\mu}
+ f^{(4)}_{a_{1}a_{2}a_{3}}~\partial_{\mu}u_{a_{1}} v_{a_{2}}^{\mu} \partial_{\nu}v_{a_{3}}^{\nu} 
\nonumber\\
+ f^{(5)}_{a_{1}a_{2}a_{3}}~\partial_{\mu}u_{a_{1}} v_{a_{2}}^{\nu} \partial^{\mu}v_{a_{3}\nu} 
+ f^{(6)}_{a_{1}a_{2}a_{3}}~\partial_{\mu}u_{a_{1}} v_{a_{2}}^{\nu} \partial_{\nu}v_{a_{3}}^{\mu} 
\nonumber\\
+ f^{(7)}_{a_{1}a_{2}a_{3}}~u_{a_{1}} v_{a_{2}}^{\mu} \partial_{\mu}\partial_{\nu}v_{a_{3}}^{\nu}
+ f^{(8)}_{a_{1}\{a_{2}a_{3}\}}~u_{a_{1}} \partial_{\mu}v_{a_{2}\nu} \partial^{\mu}v_{a_{3}}^{\nu}
\nonumber\\
+ f^{(9)}_{a_{1}\{a_{2}a_{3}\}}~u_{a_{1}} \partial_{\mu}v_{a_{2}}^{\nu} \partial_{\nu}v_{a_{3}}^{\mu}
+ f^{(10)}_{[a_{1}a_{2}]a_{3}}~u_{a_{1}} u_{a_{2}} \tilde{u}_{a_{3}}
\nonumber\\
+ f^{(11)}_{a_{1}a_{2}a_{3}}~u_{a_{1}} \partial_{\mu}u_{a_{2}} \partial^{\mu}\tilde{u}_{a_{3}}
+ f^{(12)}_{a_{1}a_{2}\{a_{3}a_{4}\}}~\partial_{\mu}u_{a_{1}} v_{a_{2}}^{\mu} v_{a_{3}}^{\nu}  v_{a_{4}\nu} 
\nonumber\\
+ f^{(13)}_{a_{1}\{a_{2}a_{3}\}a_{4}}~u_{a_{1}} v_{a_{2}}^{\mu} v_{a_{3}\mu} \partial_{\nu}v_{a_{4}\nu}
+ f^{(14)}_{a_{1}a_{2}a_{3}a_{4}}~u_{a_{1}} v_{a_{2}}^{\mu} v_{a_{3}}^{\nu} \partial_{\mu}v_{a_{4}\nu}
\nonumber\\
+ f^{(15)}_{a_{1}\{a_{2}a_{3}\}\{a_{4}a_{5}\}}~u_{a_{1}} v_{a_{2}}^{\mu} v_{a_{3}\mu} v_{a_{4}}^{\nu} v_{a_{5}\nu}
+ f^{(16)}_{[a_{1}a_{2}a_{3}][a_{4}a_{5}]}~u_{a_{1}} u_{a_{2}} u_{a_{3}} \tilde{u}_{a_{4}} \tilde{u}_{a_{5}}.
\label{W}
\eea
The Wess-Zumino consistency relation
\be
sA(T(x_{1},\dots,T(x_{n})) = 0
\ee
writes as
\be
d_{Q}A(T(x_{1},\dots,T(x_{n})) = 0
\ee
and is equivalent to
\be
d_{Q}W = 0.
\ee
This gives the following system of equations:
\bea
f^{(1)}_{a_{1}a_{2}} = a_{1} \leftrightarrow a_{2}
\nonumber\\
f^{(3)}_{a_{1}a_{2}a_{3}} = 0
\nonumber\\
f^{(4)}_{a_{1}a_{2}} = a_{1} \leftrightarrow a_{2}
\nonumber\\
f^{(5)}_{a_{1}a_{2}a_{3}} - f^{(6)}_{a_{2}a_{1}a_{3}} = 0
\nonumber\\
f^{(5)}_{a_{1}a_{2}a_{3}} + f^{(6)}_{a_{1}a_{2}a_{3}} = 0
\nonumber\\
f^{(7)}_{a_{1}a_{2}a_{3}} + f^{(11)}_{a_{1}a_{2}a_{3}} = 0
\nonumber\\
f^{(8)}_{a_{1}a_{2}a_{3}} + f^{(9)}_{a_{1}a_{2}a_{3}} = 0
\nonumber\\
f^{(10)}_{a_{1}a_{2}a_{3}} = 0
\nonumber\\
f^{(12)}_{a_{1}a_{2}a_{3}a_{4}} = a_{1} \leftrightarrow a_{2}
\nonumber\\
f^{(12)}_{a_{1}a_{2}a_{3}a_{4}} = a_{1} \leftrightarrow a_{3}
\nonumber\\
f^{(13)}_{a_{1}a_{2}a_{3}a_{4}} = 0
\nonumber\\
f^{(14)}_{a_{1}a_{2}a_{3}a_{4}} = 0
\nonumber\\
f^{(14)}_{a_{1}a_{2}a_{3}a_{4}} = 0
\nonumber\\
f^{(14)}_{a_{1}a_{2}a_{3}a_{4}} = a_{2} \leftrightarrow a_{3}
\nonumber\\
f^{(15)}_{a_{1}a_{2}a_{3}a_{4}a_{5}} = 0
\nonumber\\
f^{(16)}_{a_{1}a_{2}a_{3}a_{4}a_{5}} = 0
\eea
where, as before we have considered the coefficients of the Wick monomials
\bea
\partial_{\mu}u_{a_{1}} \partial^{\mu}u_{a_{2}}, u_{a_{1}} \partial_{\mu}u_{a_{2}} v_{a_{3}}^{\mu}, 
\partial_{\mu}u_{a_{1}} \partial^{\mu}u_{a_{2}} \partial_{\nu}v_{a_{3}}^{\nu}, 
\partial_{\mu}u_{a_{1}} \partial_{\nu}u_{a_{2}} \partial^{\mu}v_{a_{3}}^{\nu}, 
\partial^{\mu}u_{a_{1}} v_{a_{2}}^{\nu} \partial_{\mu}\partial_{\nu}v_{a_{3}}^{\nu},
\nonumber\\
u_{a_{1}} \partial^{\mu}u_{a_{2}} \partial_{\mu}\partial_{\nu}v_{a_{3}}^{\nu}, 
u_{a_{1}} \partial_{\mu}\partial_{\nu}u_{a_{2}} \partial^{\mu}v_{a_{3}}^{\nu}, 
u_{a_{1}} u_{a_{2}} \partial_{\mu}v_{a_{3}}^{\mu}, 
\partial_{\mu}u_{a_{1}} \partial^{\mu}u_{a_{2}} v_{a_{3}}^{\nu} v_{a_{4}\nu},
\partial_{\mu}u_{a_{1}} \partial_{\nu}u_{a_{2}} v_{a_{3}}^{\mu} v_{a_{4}}^{\nu},
\nonumber\\
u_{a_{1}} \partial_{\mu}u_{a_{2}} v_{a_{3}}^{\mu} \partial_{\nu}v_{a_{4}}^{\nu},
u_{a_{1}} \partial^{\mu}u_{a_{2}} v_{a_{3}}^{\nu} \partial_{\mu}v_{a_{4}\nu},
u_{a_{1}} \partial_{\mu}u_{a_{2}} v_{a_{3}}^{\nu} \partial_{\nu}v_{a_{4}}^{\mu},
u_{a_{1}} \partial_{\mu}\partial_{\nu}u_{a_{2}} v_{a_{3}}^{\mu} v_{a_{4}}^{\nu},
\nonumber\\
u_{a_{1}} \partial_{\mu}u_{a_{2}} v_{a_{3}}^{\mu} v_{a_{4}}^{\nu} v_{a_{5}\nu},
u_{a_{1}} u_{a_{2}} u_{a_{3}} \tilde{u}_{a_{4}} \partial_{\mu}v_{a_{5}}^{\mu}.
\eea
The solution of the system is:
\bea
f^{(j)}_{\dots} = 0, \quad j = 3, 10, 13, 14, 15, 16
\nonumber\\
f^{(6)}_{a_{1}a_{2}a_{3}} = f^{(5)}_{a_{2}a_{1}a_{3}}
\nonumber\\
f^{(11)}_{a_{1}a_{2}a_{3}} = - f^{(7)}_{a_{1}a_{2}a_{3}}
\nonumber\\
f^{(9)}_{a_{1}a_{2}a_{3}} = - f^{(8)}_{a_{1}a_{2}a_{3}}
\nonumber\\
f^{(5)}_{a_{1}a_{2}a_{3}} = - (a_{1} \leftrightarrow a_{2})
\nonumber\\
f^{(12)}_{a_{1}a_{2}a_{3}a_{4}} = f^{(12)}_{\{a_{1}a_{2}a_{3}a_{4}\}}
\eea
so the Wick polynomial from the expression of the anomaly simplifies to:
\bea
W= f^{(1)}_{a_{1}a_{2}}~\partial_{\mu}u_{a_{1}} v_{a_{2}}^{\mu}
+ f^{(2)}_{a_{1}a_{2}} u_{a_{1}} \partial_{\mu}v_{a_{2}}^{\mu}
\nonumber\\
+ f^{(4)}_{a_{1}a_{2}a_{3}}~\partial_{\mu}u_{a_{1}} v_{a_{2}}^{\mu} \partial_{\nu}v_{a_{3}}^{\nu} 
+ f^{(5)}_{a_{1}a_{2}a_{3}}~(\partial_{\mu}u_{a_{1}} v_{a_{2}}^{\nu} \partial^{\mu}v_{a_{3}\nu} 
+ \partial_{\mu}u_{a_{2}} v_{a_{1}\nu} \partial^{\nu}v_{a_{3}}^{\mu} )
\nonumber\\
+ f^{(7)}_{a_{1}a_{2}a_{3}}~( u_{a_{1}} v_{a_{2}}^{\mu} \partial_{\mu}\partial_{\nu}v_{a_{3}}^{\nu}
- u_{a_{1}} \partial^{\mu}u_{a_{2}} \partial_{\mu}\tilde{u}_{a_{3}} )
\nonumber\\
+ f^{(8)}_{a_{1}\{a_{2}a_{3}\}}~(u_{a_{1}} \partial_{\mu}v_{a_{2}\nu} \partial^{\mu}v_{a_{3}}^{\nu}
- u_{a_{1}} \partial_{\mu}v_{a_{2}\nu} \partial^{\nu}v_{a_{3}}^{\mu} )
\nonumber\\
+ f^{(12)}_{a_{1}a_{2}\{a_{3}a_{4}\}}~\partial_{\mu}u_{a_{1}} v_{a_{2}}^{\mu} v_{a_{3}}^{\nu}  v_{a_{4}\nu}
\eea

The previous expression can be rewritten as:
\bea
W= - {i \over 2}~f^{(1)}_{a_{1}a_{2}}~d_{Q} (v_{a_{1}\mu} v_{a_{2}}^{\mu} )
+ i~f^{(2)}_{a_{1}a_{2}}d_{Q} ( u_{a_{1}} \tilde{u}_{a_{2}} )
\nonumber\\
- {i \over 2}~f^{(4)}_{a_{1}a_{2}a_{3}}~d_{Q} (v_{a_{1}\mu} v_{a_{2}}^{\mu} \partial_{\nu}v_{a_{3}}^{\nu} )
 - {i \over 2}~f^{(5)}_{a_{1}a_{2}a_{3}}~d_{Q} (v_{a_{1}\mu} v_{a_{2}\nu} F_{a_{3}}^{\mu\nu} ) 
\nonumber\\
 - i~f^{(7)}_{a_{1}a_{2}a_{3}}~d_{Q} ( u_{a_{1}} v_{a_{2}}^{\mu} \partial_{\mu}\tilde{u}_{a_{3}} )
\nonumber\\
+ {1\over 2}~f^{(8)}_{a_{1}a_{2}a_{3}}~u_{a_{1}} F_{a_{2}\mu\nu} F_{a_{3}}^{\mu\nu}
\nonumber\\
- {i \over 4}~ f^{(12)}_{a_{1}a_{2}a_{3}a_{4}}~d_{Q} (v_{a_{1}\mu} v_{a_{2}}^{\mu} v_{a_{3}}^{\nu}  v_{a_{4}\nu} )
\eea
so, if we redefine the chronological product
$
T(T(x_{1}),\dots,T(x_{n}))
$
the anomaly is determined by 
\be
W = {1\over 2}~f_{a_{1}\{a_{2}a_{3}\}}~u_{a_{1}} F_{a_{2}\mu\nu} F_{a_{3}}^{\mu\nu}
\ee
The 
$su(2)$-invariance of the tensor
$
f_{a_{1}\{a_{2}a_{3}\}}
$
gives in fact
$
f_{a_{1}\{a_{2}a_{3}\}} = 0
$
because of the symmetry property in the last two indexes. So we end up with
\be
W = 0\quad  \Longrightarrow \quad A(T(x_{1},\dots,T(x_{n})) = 0.
\ee
This finishes the induction.
$\qed$

\begin{rem}
We might be tempted to use the identity
\be
\square f_{j} = 0, j = 1, 2, 3 \quad \Longrightarrow
\partial_{\mu}f_{1} \partial^{\mu}f_{2} f_{3} = {1\over 2} ( \partial^{\mu}f_{1} f_{2} f_{3} + f_{1} \partial^{\mu}f_{2} f_{3}
- f_{1} f_{2} \partial^{\mu}f_{3}) 
\ee
to fix:
\be
f^{(j)}_{\dots} = 0, \quad j = 8, 11
\ee
in the expression (\ref{W}) of the polynomial $W$. However, to do that we must redefine the chronological products
$
T(T^{\mu}(x_{1}), \dots)
$
and this is forbidden: we had to fix these chronological products at the step (iv).
\end{rem}

\begin{rem}
Apparently our result is weaker than the result from \cite{FHH} where the annulment of the anomaly is proved for 
all simple compact Lie algebras. Our result extends only to Lie algebras for which there are no invariant tensors of the type
$
f_{a_{1}\{a_{2}a_{3}\}}
$.
\end{rem}

\begin{rem}
In the case of massive Yang-Mills fields of equal mass
$m$, one has to add:
(a) ghost scalar fields
$
\Phi_{a}
$
of the same mass $m$ to be able to describe particles of spin $1$ and mass $m$;
(b) a (physical) scalar field (the Higgs field) 
$\phi$
of mass
$
m_{H}
$
to have gauge invariance in the second order of perturbation theory \cite{DS}. However, in this case the method from the 
previous theorem does not work. In the last step of the proof, we are left with a contribution
\be
W = \lambda_{1} u_{a} \Phi_{a} \Phi_{b} \Phi_{b} \phi + \lambda_{2} u_{a} \Phi_{a} \phi^{3}
\ee
which cannot be eliminated by a redefinition of the chronological products. Again, we remark that using the flow equation
method, one can prove gauge invariance \cite{KM}.
\end{rem}
\newpage
\section{Quantum Electrodynamics\label{qed}}

The precedent method is quite effective in analyzing simpler models like quantum electrodynamics (QED). We present the 
analysis from \cite{ano} in a simplified form. The fields are: the photon field described as in section \ref{free} by
$
v_{\mu}, u, \tilde{u}
$
and the Dirac field
$
\psi, \bar{\psi}
$
of mass $m$ determined by the Fermi statistics and the two-point function
\bea
<\Omega, \psi_{\alpha}(x_{1}) \bar{\psi}_{\beta}(x_{2})\Omega> = 
- i~S_{m}^{(+)}(x_{1} - x_{2})_{\alpha\beta}
\nonumber \\
<\Omega, \bar{\psi}_{\alpha}(x_{1}) \psi_{\beta}(x_{2})\Omega> = 
- i~S_{m}^{(-)}(x_{2} - x_{1})_{\beta\alpha}
\label{2-dirac}
\eea
where
\be
S_{m}^{(\epsilon)}(x) \equiv (i~\gamma^{\mu}~\partial_{\mu} + m)~D_{m}^{(\epsilon)}(x), \quad \epsilon = \pm.
\label{S}
\ee

The interaction Lagrangian is then
\be
T = v_{\mu}~\bar{\psi}~\gamma^{\mu}~\psi
\ee
and one can find that the procedure (\ref{descent}) stops at:
\be
T^{\mu} = u~\bar{\psi}~\gamma^{\mu}~\psi. 
\ee

As in the previous section, we can take the anomalies of the form
\be
A(T^{I_{1}}(x_{1}),\dots,T^{I_{n}}(x_{n})) =  \delta(X)~W^{I_{1},\dots,I_{n}}(x_{1})
\ee
with 
$
W^{\dots}
$
some Wick polynomials (constrained by canonical dimension, ghost number and Lorentz covariance - as before).

Now, because we have only one ghost field of ghost number $1$, namely $u$ it follows easily that we must have:
\be
A(T^{I_{1}}(x_{1}),\dots,T^{I_{n}}(x_{n})) =  0, \quad |I_{1}| + \cdots |I_{n}| \geq 2.
\ee

Next, we immediately have that
\be
W^{\mu} \equiv W^{[\mu],\emptyset,\dots,\emptyset}
\ee
must be of the form:
\be
W^{\mu} = a_{1} u \partial^{\mu}u + a_{2} u \partial^{\mu}u v_{\rho} v^{\rho} + a_{3} u \partial_{\rho}u v_{\rho} v^{\mu}
\ee
for some constants
$a_{i}$.

The Wess-Zumino consistency condition 
\be
d_{Q}A(T^{\mu}(x_{1}), \dots) = 0 \quad \Rightarrow \quad  d_{Q}~W^{\mu} = 0
\ee
gives
\be
a_{3} = 2 a_{1}.
\ee

It follows that 
\be
W^{\mu} = i d_{Q} (a_{1} u v^{\mu} + a_{2} u v^{\mu} v_{\rho} v^{\rho})
\ee
so, if we perform the finite renormalization
\be
T(T^{\mu}(x_{1}),T(x_{2}),\dots,T(x_{n})) \rightarrow
T(T^{\mu}(x_{1}),T(x_{2}),\dots,T(x_{n})) 
+ i \delta(X) U_{2}^{\mu}(x_{1})
\label{R3}
\ee
we make:
\be
A(T^{I_{1}}(x_{1}),\dots,T^{I_{n}}(x_{n})) =  0, \quad |I_{1}| + \cdots |I_{n}| = 1.
\ee

Finally we determine the generic form of
\be
W \equiv W^{\emptyset,\dots,\emptyset}
\ee
as:
\bea
W = a_{1} u + a_{2} \partial_{\mu}u v^{\mu} + a_{3} u \partial_{\mu}v^{\mu}
+ a_{4} \partial_{\mu}u \partial^{\mu}\partial^{\nu}v_{\nu} 
+ a_{5} \partial_{\mu}\partial_{\nu}u \partial^{\mu}v^{\nu}
\nonumber\\
+ a_{6} u v_{\mu} v^{\mu} + a_{7} u v^{\mu} \partial_{\mu}\partial_{\nu}v^{\nu}
+ a_{8} u \partial^{\mu}v^{\nu} \partial_{\mu}v_{\nu}
+ a_{9} u \partial^{\mu}v^{\nu} \partial_{\nu}v_{\mu}
+ a_{10} \partial_{\mu}u v^{\mu} \partial^{\nu}v_{\nu} 
\nonumber\\
+ a_{11} \partial_{\mu}u v_{\nu} \partial^{\mu}v^{\nu}
+ a_{12} \partial_{\mu}u v^{\mu} \partial^{\nu}v^{\mu} 
+ a_{13} \partial_{\mu}\partial_{\nu}u v^{\mu} v^{\nu}
\nonumber\\
+ a_{14} u v^{\mu} v^{\nu} \partial_{\mu}v_{\nu}
+ a_{15} \partial_{\mu}u v^{\mu} v^{\nu} v_{\nu}
+ a_{16} u v^{\mu} v_{\mu} v^{\nu} v_{\nu}
\nonumber\\
+ b_{1} u \bar{\psi} \psi
+ b_{2} u v_{\mu} \bar{\psi} \gamma^{\mu} \psi
\nonumber\\
+ a^{\prime} u \epsilon_{\mu\nu\rho\sigma} F^{\mu\nu} F^{\rho\sigma}
+ b^{\prime} u \bar{\psi} \gamma_{5} \psi
\label{W1}
\eea
The expressions 
$
\sim a_{j}, \quad j = 2, \dots, 5, 10, 15
$
are coboundaries, i.e.
$
\sim d_{Q}U_{1}.
$
The Wess-Zumino consistency condition 
\be
d_{Q}A(T(x_{1}), \dots) = 0 \quad \Rightarrow \quad  d_{Q}~W = 0
\ee
gives
\bea
a_{j} = 0, \quad j = 6, 7, 14, 16
\nonumber\\
a_{9} = - a_{8}
\nonumber\\
a_{12} = a_{11}
\nonumber\\
a_{13} = 2 a_{11}
\eea
and it follows that the expressions
$
\sim a_{11}
$
is a coboundaries, i.e.
$
\sim d_{Q}U_{2}.
$
In the end we are left with
\be
W = a_{1} u + {1\over 4} a_{8} u F^{\mu\nu} F_{\mu\nu} +  b_{1} u \bar{\psi} \psi 
+ a^{\prime} u \epsilon_{\mu\nu\rho\sigma} F^{\mu\nu} F^{\rho\sigma} + b^{\prime} u \bar{\psi} \gamma_{5} \psi
+ d_{Q} U.
\ee
If we impose charge conjugation invariance \cite{Sc2} the first five terms must be zero. 
The last term can be eliminated by a redefinition of the chronological product
$
T(T(x_{1}),\dots,T(x_{n}))
$
so we have 
\be
A(T(x_{1}),\dots,T(x_{n})) = 0.
\ee

\section{Conclusions}

We would to understand better why the method of \cite{FHH} seems to be stronger. If some supplementary condition is imposed,
this condition should be translated in terms of chronological products and its physical meaning should be investigated.


\newpage
\begin{thebibliography}{99}

\bibitem{BS}
N. N. Bogoliubov, D. Shirkov,
``{\it Introduction to the Theory of Quantized Fields}",
John Wiley and Sons, 1976 (3rd edition)

\bibitem{BLOT}
N. N. Bogoliubov, A. A. Logunov, A.I. Oksak, I. Todorov, 
``{\it General Principles of Quantum Field Theory}", Kluwer 1989

\bibitem{DF}
M. Duetsch, K. Fredenhagen,
``{\it A Local (Perturbative) Construction of Observables in Gauge Theories:
the Example of QED}",\\
arXiv:9807078, Commun. Math. Phys. {\bf 203} (1999) 71-105

\bibitem{DS}
M. Duetsch, Bert Schroer, ``{\it Massive vector mesons and gauge theory}", \\arXiv: 9906089, 
J. Phys. A: Math. Gen. {\bf 33} (2000) 4317 - 4356

\bibitem{EG}
H. Epstein, V. Glaser,
``{\it The R\^ole of Locality in Perturbation Theory}",\\
Ann. Inst. H. Poincar\'e {\bf 19 A} (1973) 211-295

\bibitem{FHH}
M. B. Fr\"ob, J. Holland, S. Hollands,
``{\it All-order bounds for correlation functions of gauge-invariant operators in Yang - Mills theory}",\\
arXiv:1511.09425v2, J. Math. Phys. {\bf 57} (2016) no.12, 122301

\bibitem{Gl}
V. Glaser,
``{\it Electrodynamique Quantique}",
L'enseignement du 3e cycle de la physique en Suisse Romande (CICP), Semestre
d'hiver 1972/73

\bibitem{standard} D. R. Grigore,
``{\it The Standard Model and its Generalisations in Epstein-Glaser Approach to
Renormalisation Theory}",\\
arXiv:9810078, Journ. Phys. {\bf A 33} (2000) 8443-8476 

\bibitem{ano} D. R. Grigore,
``{\it The Structure of the Anomalies of Gauge Theories in the Causal Approach}", \\
arXiv:0010226, Journ. Phys. {\bf A 35} (2002) 1665-1689

\bibitem{cohomology}
D. R. Grigore, 
``{\it Cohomological Aspects of Gauge Invariance in the Causal Approach}",\\
Romanian Journ. Phys. {\bf 55} (2010) 386-438
 
\bibitem{off}
D. R. Grigore,
``{\it Off-Shell Fields and Quantum Anomalies}",\\
arXiv:1011.3219, (conf. C\u aciulata, 2010), Physics Annals of the
University of Craiova, PAUC, vol. 21 - special issue (2011) 117-130

\bibitem{Ha}
R. Haag ``{\it Local Quantum Physics: Fields, Particles, Algebras}", second edition. Springer, 1992

\bibitem{H}
K. Hepp, ``{\it Renormalization Theory}", in ``{\it Statistical Mechanics and Quantum Field Theory}" pp. 429 - 500,
(Les Houches 1970), C. DeWitt-Morette, Raymond Stora (eds.), Gordon and Breach 1971

\bibitem{HT}
M. Henneaux, C. Teitelboim, ``{\it  Quantization of Gauge Systems}" Princeton Univ. Press, 1992

\bibitem{J}
R. Jost, 
``{\it The General Theory of Quantized Fields}", 
AMS, Providence, 1965

\bibitem{KM}
C. Kopper, V. F. Muller, ``{\it Renormalization of Spontaneously Broken $SU(2)$ Yang-Mills Theory with Flow Equations}",\\
arXiv:1704.06799v2, Rev. Math. Phys. {\bf 21} (2009) 781 - 820

\bibitem{P}
J. Polchinski, ``{\it Renormalization and Effective Lagrangians}",\\
Nucl. Phys. {\it B 231} (1984) 269 - 295

\bibitem{PS}
G. Popineau, R. Stora, ``{\it A Pedagogical Remark on the Main Theorem of Perturbative Renormalization Theory}", \\
Nuclear Physics {\bf B 912} (2016) 70 - 78

\bibitem{S}
M. Salmhofer, ``{\it Renormalization: An Introduction}", (Theoretical and Mathematical Physics) Springer 1999

\bibitem{Sc2}
G. Scharf, ``{\it Finite Quantum Electrodynamics: The Causal Approach}", third edition, Dover, 2014

\bibitem{Sto1}
R. Stora,
``{\it Lagrangian Field Theory}",
Les Houches lectures, Gordon and Breach, N.Y., 1971, 
C. De Witt, C. Itzykson eds.

\bibitem{St1}
O. Steinmann,
``{\it Perturbation Expansions in Axiomatic Field Theory}",
Lect. Notes in Phys. {\bf 11}, Springer, 1971

\bibitem{SW}
R. F. Streater, A. S. Wightman,
``{\it PCT, Spin and Statistics and all that}", 
W. A. Benjamin Inc. New York, 1964

\bibitem{Z-J}
J. Zinn-Justin, ``{\it Renormalization of Gauge Theories}", in
``{\it Trends in Elementary Particle Theory}", (International Summer Institute on Theoretical Physics in Bonn 1974), pp. 2 - 39,
H. Rollnik, K. Dietz (eds.), Springer 1975
\end{thebibliography}
\end{document}